\documentclass[a4paper,11pt]{article}

\usepackage{jcappub} 

\usepackage[T1]{fontenc} 
\RequirePackage{bm,amsfonts,lineno,amsmath,amssymb,float,eurosym,latexsym,epsf,mathtools,cuted,makecell,array}
\usepackage[font={footnotesize,it}]{caption}
\setcellgapes{4pt}

\newcommand*\DAlambert{\mathop{}\!\mathbin\Box}

\title{\boldmath Strange stars in $f(R,\mathcal{T})$ gravity}

\author[a]{Debabrata Deb,}
\author[b]{Farook Rahaman,}
\author[c,1]{Saibal Ray\note{Corresponding author.},}
\author[a]{B. K. Guha}

\affiliation[a]{Department of Physics, Indian Institute of Engineering Science and Technology, Shibpur, Howrah 711103, West Bengal, India}
\affiliation[b]{Department of Mathematics, Jadavpur University, Kolkata 700032, West Bengal, India}
\affiliation[c]{Department of Physics, Government College of Engineering and Ceramic Technology, Kolkata 700010, West Bengal, India}

\emailAdd{ddeb.rs2016@physics.iiests.co.in}
\emailAdd{rahaman@associates.iucaa.in}
\emailAdd{saibal@associates.iucaa.in}
\emailAdd{bkguhaphys@gmail.com}

\abstract{In this article we try to present spherically symmetric isotropic strange star model under the framework of  $f(R,\mathcal{T})$ theory of gravity. To this end, we consider that the Lagrangian density is a linear function of the Ricci scalar $R$ and the trace of the energy momentum tensor~$\mathcal{T}$ given as $f\left(R,\mathcal{T}\right)=R+2\chi T$. We also assume that the quark matter distribution is governed by the simplest form of the MIT bag model equation of state (EOS) as $p=\frac{1}{3}\left(\rho-4B\right)$, where $B$ is the bag constant. We have obtained an exact solution of the modified form of the Tolman-Oppenheimer-Volkoff (TOV) equation in the framework of $f(R,\mathcal{T})$ gravity theory and have studied the dependence of different physical properties, viz., the total mass, radius, energy density and pressure for the chosen values of $\chi$. Further, to examine physical acceptability of the proposed stellar model, we have conducted different tests in detail, viz., the energy conditions, modified TOV equation, mass-radius relation, causality condition etc. We have precisely explained the effects arising due to the coupling of the matter and geometry on the compact stellar system. For a chosen value of the bag constant, we have predicted numerical values of the different physical parameters in tabular form for the different strange star candidates. It is found that as the factor $\chi$ decreases the strange star candidates become gradually massive and larger in size with less dense stellar configuration. However, when $\chi$ increases the stars shrink gradually and become less massive to turn into a more compact stellar system. Hence for $\chi>0$ our proposed model is suitable to explain the ultra-dense compact stars well within the observational limits and for $\chi<0$ case allows to represent the recent massive pulsars and super-Chandrasekhar stars. For $\chi=0$ we retrieve as usual the standard results of the general relativity (GR).}


\begin{document}
\maketitle
\flushbottom

\section{Introduction}\label{sec:intro}\label{sec:intro}
We are in the new avenue of the modern cosmology immediately after the observational evidence~\cite{Riess1998,Perlmutter1999,Bernardis2000,Perlmutter2003} of late-time acceleration
of the Universe. This invites a serious theoretical challenge to Einstein's general theory of relativity, which was proven unquestionable for the last several decades. To address this situation in the framework of GR many authors~\cite{Caldwell2002,Nojiri2003,Odinstov2003,Padmanabhan2002,Kamenshchik2001,Bento2002} introduced presence of exotic entity, widely known as dark energy. On the other hand, a class of authors proposed different modified theories time and again to address the present accelerating phase of the universe. Essentially by replacing the standard Einstein-Hilbert action by arbitrary function of Ricci scalar $R$, i.e., $f(R)$ it was  successfully possible to explain the late time cosmic acceleration, unified inflation with dark energy and also galactic dynamics of massive test particles without introducing any exotic dark energy~\cite{Carroll2004,Nojiri2007,Nojiri2008,Cognola2008,Cognola2011,Nojiri2011}. For the detailed work on $f(R)$ gravity one may consult the following literature~\cite{Lobo2008,Sotiriou2010,Capozziello2010}.

Recently, Harko et al.~\cite{harko2011} have presented more generalized form of the $f(R)$ gravity theory by choosing the Lagrangian density as the arbitrary function $f (R, \mathcal{T} )$ where $R$ is the Ricci scalar and $\mathcal{T}$ is the trace of the energy-momentum tensor. This is known as $f (R, \mathcal{T} )$ theory of gravity and depends on a source term, which represents the variation of the energy-momentum tensor with respect to the metric, due to coupling of the matter and geometry. Here, for the presence of an extra force perpendicular to the four velocity, the test particles do not follow the geodesic path. However, lateron Chakraborty~\cite{SC2013} showed that due to the linear form of the function $f(R,\mathcal{T})$ given as $f(R,\mathcal{T})=R+h(T)$, the whole system behaves like a non-interacting 2-fluid system and test particles follow the geodesic path. The interaction between the matter and curvature terms can be interpreted as the origin of the 2nd type of fluid. In his detailed work Chakraborty~\cite{SC2013} also explained how rapidly the 2nd type of fluid becomes exotic though the first type is still a normal fluid. 

In this line of thinking many literature can be referred~\cite{Harko-2008,Bisabr-2012,Jamil12,Alv13,shabani13,shabani14,Zare,shabani16a,shabani16b} where $f(R,\mathcal{T})$ theory of gravity have been used successfully. Sharif et al.~\cite{sharif2014} indicated the factors which are effecting the stability of the isotropic and spherically symmetric stellar system in $f(R,\mathcal{T})$ gravity. To find the collapse equation for the Newtonian and post-Newtonian eras by employing a perturbation scheme on the dynamical equations Noureen et al.~\cite{noureen2015} imposed condition on adiabatic index. Further, by employing the perturbation method on the modified field equations Noureen and Zubair~\cite{zubair2015a,noureen2015b,noureen2015c} in a series of papers studied dynamical analysis of a spherically symmetric collapsing star under different physical situations. Ahmed et al.~\cite{Ahmed2015} studied the effect of $f(R,\mathcal{T})$ gravity on gravitational lensing and compared the yielded result with the standard result of GR. Using Runge-Kutta 4th-order method to solve the TOV equation in $f(R,\mathcal{T})$ gravity Moraes et al.~\cite{Moraes2015} studied hydrostatic equilibrium configurations for the neutron stars and strange stars. Later on Das et al.~\cite{Amit2016} used the results to study spherically symmetric and isotropic compact stellar system in $f(R,\mathcal{T})$ gravity by adopting the Lie algebra with conformal Killing vectors. Under the $f(R,\mathcal{T})$ gravity Das et al.~\cite{Amit2017} also presented a model for Gravastar to avoid singularity and thus provided an alternative to black hole.

In this article, keeping the above works in mind, we have presented an exact solution of the TOV equation for the spherically symmetric and isotropic strange star candidates in the framework of $f(R,\mathcal{T})$ theory of gravity. The outline of the present study is organized as follows: In section~\ref{sec1} we present basic mathematical formulation of $f(R,\mathcal{T})$ theory. The Einstein field equations are shown in section~\ref{sec2}. Solutions of the field equations for $f\left(R,\mathcal{T}\right)$ gravity are obtained in section~\ref{sec3}. In section~\ref{sec4} we discuss some physical properties of our stellar system, such as the energy conditions~\ref{subsec4.1}, mass-radius relation~\ref{subsec4.2}, compactification factor and redshift~\ref{subsec4.3} along with the stability conditions~\ref{subsec4.4}, viz., modified TOV equation~\ref{subsubsec4.4.1}, causality condition~\ref{subsubsec4.4.2} and adiabatic index~\ref{subsubsec4.4.3}. Finally, in section~\ref{sec5} we make some concluding remarks.

\section{Basic mathematical formulation of $f(R,\mathcal{T})$ Theory}\label{sec1}
Following Harko et al.~\cite{harko2011}, we propose the action due to $f(R,\mathcal{T})$ gravity which reads
\begin{equation}\label{1.1}
S=\frac{1}{16\pi}\int d^{4}xf(R,\mathcal{T})\sqrt{-g}+\int
d^{4}x\mathcal{L}_m\sqrt{-g},
\end{equation}
where $f(R,\mathcal{T})$, $g$ and $\mathcal{L}_m$ are the general function of the Ricci scalar $R$
and the trace of the energy-momentum tensor $\mathcal{T}$, the determinant of the metric $g_{\mu\nu}$ 
and the matter Lagrangian density, respectively. Further we consider the geometrical units $G=c=1$. 
 
Now variation of the action (\ref{1.1}) with respect to $g_{\mu\nu}$ 
yields the field equations of the $f(R,\mathcal{T})$ gravity model as follows:
\begin{eqnarray}\label{1.2}
&\qquad\hspace{-3cm} f_R (R,\mathcal{T}) R_{\mu\nu} - \frac{1}{2} f(R,\mathcal{T}) g_{\mu\nu} 
+ (g_{\mu\nu}\DAlambert - \nabla_{\mu} \nabla_{\nu}) f_R (R,\mathcal{T})\nonumber \\
&\qquad\hspace{5cm} = 8\pi T_{\mu\nu} - f_\mathcal{T}(R,\mathcal{T}) T_{\mu\nu} -
f_\mathcal{T}(R,\mathcal{T})\Theta_{\mu\nu},
\end{eqnarray}
where {{$f_R (R,\mathcal{T})= \partial f(R,\mathcal{T})/\partial R$,
$f_\mathcal{T}(R,\mathcal{T})=\partial f(R,\mathcal{T})/\partial
\mathcal{T}$. Again, $\DAlambert \equiv\partial_{\mu}(\sqrt{-g} g^{\mu\nu} \partial_{\nu})/\sqrt{-g}$}} is the D'Alambert operator, $R_{\mu\nu}$ is the Ricci tensor, $\nabla_\mu$ represents the
covariant derivative associated with the Levi-Civita connection of $g_{\mu\nu}$, $\Theta_{\mu\nu}=
g^{\alpha\beta}\delta T_{\alpha\beta}/\delta g^{\mu\nu}$ and the
stress-energy tensor defined as $T_{\mu\nu}=g_{\mu\nu}\mathcal{L}_m-2\partial\mathcal{L}_m/\partial
g^{\mu\nu}$~\cite{Landau2002}.

The covariant divergence of eq. (\ref{1.2}) gives~\cite{barrientos2014}
\begin{eqnarray}\label{1.3}
\hspace{-0.5cm}\nabla^{\mu}T_{\mu\nu}&=&\frac{f_\mathcal{T}(R,\mathcal{T})}{8\pi -f_\mathcal{T}(R,\mathcal{T})}[(T_{\mu\nu}+\Theta_{\mu\nu})\nabla^{\mu}\ln f_\mathcal{T}(R,\mathcal{T}) +\nabla^{\mu}\Theta_{\mu\nu}-\frac{1}{2}g_{\mu\nu}\nabla^{\mu}\mathcal{T}].
\end{eqnarray}

The above eq. (\ref{1.3}) features that the energy-momentum tensor is not conserved from the point of view of the General Relativity (GR). Basically in the covariant derivative of the stress-energy tensor as $f_\mathcal{T}(R,\mathcal{T}) \neq 0$ (for our model $f_\mathcal{T}(R,\mathcal{T})=2\chi$) and hence $\nabla^{\mu}T_{\mu\nu}\neq 0$. As a result, the system will not be conserved as usual in Einsteinian gravity. This aspect can be shown in details with another treatment as follows:

The equation of motion for a test particle~\cite{barrientos2014} in our case can be provided as
\begin{eqnarray}\label{2}
\frac{d^2x^{\mu}}{ds^2}+\Gamma^{\mu}_{\lambda\nu}{u^{\lambda}}{u^{\nu}}=f^{\mu},
\end{eqnarray}
where $f^{\mu}$ is the effective force and is given by 
\begin{eqnarray}\label{3}
f^{\mu}=\frac{\left(8\pi{\nabla}_{\nu}p-\frac{1}{2}f_\mathcal{T}(R,\mathcal{T}){\nabla}_{\nu}\mathcal{T} \right)}{\left(\rho+p\right)\left[8\pi+f_\mathcal{T}(R,\mathcal{T})\right]} h^{\mu\nu},
\end{eqnarray}
where $h_{\mu\nu}=g_{\mu\nu}-{u_{\mu}}{u_{\nu}}$ is a projection operator.

Introducing the trace $\mathcal{T}=\rho-3p$ and for the vanishing pressure $p$ we find 
\begin{eqnarray}\label{4}
f^{\mu}=-\frac{f_\mathcal{T}(R,\mathcal{T}){\nabla}_{\nu}\rho}{2 \rho \left[8\pi+f_\mathcal{T}(R,\mathcal{T})\right]} h^{\mu\nu}.
\end{eqnarray}

Clearly unlike the usual Einsteinian gravity the particles do not follow a geodesic path. Hence, the energy-momentum tensor is not conserved in $f(R,\mathcal{T})$ gravity from the point of view of GR.
 
Now, we are considering the energy-momentum tensor of a perfect fluid given as
\begin{equation}\label{1.4}
T_{\mu\nu}=(\rho+p)u_\mu u_\nu-{p}g_{\mu\nu},
\end{equation}
where ${v_{\mu}}$ and ${u_{\nu}}$ are the radial-four vectors and four velocity vectors, respectively. Here $\rho$ and $p$ represent the matter density and pressure of the fluid, respectively. In the present study we choose $\mathcal{L}_m=-p$ and also we have $\Theta_{\mu\nu}=-2T_{\mu\nu}-p g_{\mu\nu}$. The logic behind the choice of $\mathcal{L}_m=-p$ is as follows: here we have followed the calculations as presented by Harko et al.~\cite{harko2011} where they assumed that ${\mathcal{L}}_m = -p$. Again, Faraoni~\cite{Faraoni2009} showed that both the Lagrangian ${\mathcal{L}}_m=p$ and  ${\mathcal{L}}_m= -\rho$ are equivalent when the fluid couples minimally to the gravity. Also, Harko in his study~\cite{Harko2010} showed that the Lagrangian expressed in terms of $\rho$ and $p$ only are completely equivalent, as $\rho$ and $p$ are freely interchangeable thermodynamic quantities. Thus the assumption ${\mathcal{L}}_m = -p$ used in the present work is quite arbitrary and does not change the subsequent results whatever the form one adopts.

We assume the simplified and linear functional form of $f(R,\mathcal{T})$ as suggested by Harko et al.~\cite{harko2011} which is given as $f(R,\mathcal{T})=R+2\chi\mathcal{T}$, where $\chi$ is a constant. Many authors successfully used this functional form in their $f(R,\mathcal{T})$ models~\cite{singh2015,moraes2014b,moraes2015a,moraes2015b,moraes2017,singh2014,baffou2015,shabani13,shabani14,sharif2014b,reddy2013b,kumar2015,shamir2015,Fayaz2016}.

Now substituting the specific form of $f(R,\mathcal{T})$ in eq.~(\ref{1.2}) we have the modified form of the Einstein field equation as follows
\begin{equation}\label{1.5}
G_{\mu\nu}=8\pi T_{\mu\nu}+\chi \mathcal{T}g_{\mu\nu}+2\chi(T_{\mu\nu}+p g_{\mu\nu})=8\pi T^{eff}_{\mu\nu},
\end{equation}
 where $G_{\mu\nu}$ is the usual Einstein tensor and $T^{eff}_{\mu\nu}=T_{\mu\nu}+\frac{\chi}{8\pi} \mathcal{T}g_{\mu\nu}+\frac{\chi}{4\pi}(T_{\mu\nu}+p g_{\mu\nu})$.
 
 Putting $\chi=0$ in eq.~(\ref{1.5}) the standard general relativistic results can easily be retrieved. Substituting $f(R,\mathcal{T})=R+2\chi\mathcal{T}$ in eq.~(\ref{1.3}) yields 
 \begin{equation}\label{1.6}
 \left(4\pi+\chi\right)\nabla^{\mu}T_{\mu\nu}=-\frac{1}{2}\chi\left[g_{\mu\nu}\nabla^{\mu}\mathcal{T}+2\nabla^{\mu}(p g_{\mu\nu})\right].
 \end{equation} 

 We may also represent eq.~(\ref{1.6}) as
 \begin{equation}\label{1.7}
 \nabla^{\mu}T^{eff}_{\mu\nu}=0.
 \end{equation}
 
 Again substituting $\chi=0$ in eqs.~(\ref{1.6}) and (\ref{1.7}) the standard form of the conservation of the energy momentum tensor for GR can be achieved.

 \section{The Einstein field equations for $f\left(R,\mathcal{T}\right)$ gravity}\label{sec2}
  We assume the interior spacetime of the spherically symmetric static stellar configuration which can be described by the metric given as
 \begin{equation}\label{2.1}
ds^2=e^{\nu(r)}dt^2-e^{\lambda(r)}dr^2-r^2(d\theta^2+\sin^2\theta d\phi^2),
\end{equation}
 where $\nu$ and $\lambda$ are the metric potentials and functions of the radial coordinate only.
 
 Now using eq.~(\ref{1.5}) along with eqs. (\ref{1.4}) and (\ref{2.1}) we have the explicit form of the Einstein field equations in $f\left(R,\mathcal{T}\right)$ gravity as~\cite{Moraes2015,Amit2016,Amit2017}
 {{\begin{eqnarray}\label{2.2}
 &\qquad\hspace{-1cm} {{\rm e}^{-\lambda}} \left( {\frac {\lambda^{{\prime}}}{r}}-\frac{1}{{r}^{2}}
 \right) +\frac{1}{{r}^{2}}= \left( 8\pi +3\chi \right) \rho-\chi p=8\pi{{\rho}^{eff}},\\ \label{2.3}
&\qquad\hspace{-1cm} {{\rm e}^{-\lambda}} \left( {\frac {\nu^{{\prime}}}{r}}+\frac{1}{{r}^{2}} \right) -\frac{1}{{r}^{2}}= \left( 8\pi +3\chi \right) p-\chi\rho=8\pi{{p}^{eff}}, 
 \end{eqnarray}}}
where $`\prime$' denotes the differentiation with respect to the radial coordinate. Also, the effective density and pressure ${\rho}^{eff}$ and ${p}^{eff}$ respectively, are given as 
\begin{eqnarray}\label{2.2a}
\rho^{eff}=\rho+\frac{\chi}{8\,\pi}\left( 3\rho-p \right), \\ \label{2.2b}
p^{eff}=p-\frac{\chi}{8\pi}\left( \rho-3p \right).
\end{eqnarray}
 
Considering all the three flavors of quarks are confined in a bag and are non-interacting in nature, we have assumed in the present study that the quark matter distribution is governed by the simplest form of the EOS given by
 \begin{equation}\label{2.4}
{p}+{B}={\sum_f}{p^f},
 \end{equation}
 where $B$ is the total external bag pressure which is a constant quantity under certain numerical range and $p^f$ is the individual pressure due to the quark flavors of up, down and strange quarks.  
 
 The de-confined quarks inside the bag has the total energy density $\rho$ as follows
 \begin{equation}
\rho={\sum_f}{{\rho}^f}+B, \label{2.5}
\end{equation}
 where ${{\rho}^f}=3{p^f}$ is the energy density of individual quark flavors.
 
 Using eqs.~(\ref{2.4}) and (\ref{2.5}) we have the EOS of quark matter as follows
 \begin{equation}
{p}=\frac{1}{3}({{\rho}}-4B),\label{2.6}
\end{equation} 
 which is the well known MIT bag EOS. 

It is worthwhile to mention here that without introducing the critical aspects of quantum mechanical particle physics the MIT bag EOS is used successfully in many literature~\cite{1,2,3,4,5,6,7,8} to study the strange star candidates in the framework of GR. Here the factor $1/3$ in eq. (\ref{2.6}) is arising due to the fact that the $u$, $s$ and $d$ quarks are considered to be massless. Farhi and Jaffe in their study~\cite{Jaffe1984} showed that for the massless and non-interacting quarks the strange quark matter (SQM) might be the true ground state of the strongly interacting matter distribution when $B$ is between $57~MeV/{fm}^3$ and $94~MeV/{fm}^3$. However, in the present study we have chosen value of the bag constant as $B=83~MeV/{{fm}^3}$~\cite{Rahaman2014}.
 
 Let us define mass of the spherically symmetric stellar system as 
 \begin{equation}\label{2.7}
m \left( r \right) =4\pi \int_{0}^{r}\!{{\rho}^{eff}} \left( r \right) {r}^
{2}{dr}.
\end{equation}

In order to obtain a non-singular monotonic decreasing matter density inside the stellar system, 
following Mak and Harko~\cite{MH2002}, we consider the simplified form of $\rho$ given as
 \begin{equation}\label{2.8}
\rho(r)=\rho_c\left[1-\left(1-\frac{\rho_0}{\rho_c}\right)\frac{r^{2}}{R^{2}}\right],
\end{equation}
where $\rho_c$ and $\rho_0$ are the constants and they represent maximum and minimum values of $\rho$, respectively. 
 
Now from the modified form of the Einstein field equation (\ref{1.5}) one can note that in the exterior spacetime $T_{\mu\nu}$ is zero as there is no matter in the vacuum spacetime. Hence we find from eq.~(\ref{1.5}) that $G_{\mu\nu}{\mid}_{ext}=0$ which means that there will be no change of GR solution for the exterior spacetime metric even in the $f(R,\mathcal{T})$ theory of gravity. Hence we adopt the exterior Schwarzschild metric given by
 \begin{eqnarray}\label{7}
 {ds}^2=\left(1-\frac{2M}{r}\right)dt^2- {\left(1-\frac{2M}{r}\right)}^{-1} dr^2-r^2(d\theta^2+\sin^2\theta d\phi^2).
 \end{eqnarray} 

Using eqs. (\ref{2.2a}) and (\ref{2.6})-(\ref{7}) we have the total mass of the stellar system as
 \begin{eqnarray}\label{2.10}
 M=\frac{2R}{45\pi}}{\left(72B{\pi}^{2}{R}^{2}+56B\pi {R}^{2}\chi+9B{R}^{2}{\chi}^{2}+12{\pi }^{2}{R}^{2}\rho_{{c}}+4\pi {R}^{2}\chi\rho_{{c}} \right).
 \end{eqnarray}
 
 Substituting eq.~(\ref{2.7}) into eq.~(\ref{2.2}) we have eventually 
 \begin{eqnarray}\label{2.11}
 {{\rm e}^{-\lambda \left(r \right) }}=1-{\frac {2m(r)}{r}}.
 \end{eqnarray}
 
Based on the above mathematical formalism, in the following Section 4, we shall solve the field equations to find out different physical parameters of the system.

\section{Solution of Einstein field equations for $f\left(R,\mathcal{T}\right)$ gravity}\label{sec3}
Now solving the Einstein field equations~(\ref{2.2}) and (\ref{2.3}) and using the eqs. (\ref{2.2a}), (\ref{2.2b}), (\ref{2.6}) and (\ref{2.10}) we obtain different physical parameters of the system given as
\begin{eqnarray}\label{3.1}
 & \qquad\hspace{-1cm} \lambda \left( r \right) =-\ln  \left[ {\frac { \left( 16\lambda_{{1
}}{r}^{2}+\pi  \right) {R}^{5}-16\,\lambda_{{1}}{r}^{4}{R}^{3}-5\,M
\pi {R}^{2}{r}^{2}+3\,M\pi {r}^{4}}{{R}^{5}\pi }} \right],\\\label{3.2}
& \qquad\hspace{-1cm} \nu \left( r \right) =\frac {1}{ \left( 12\pi +4\chi \right) \nu_
{{1}}} \Bigg[-8\nu_{{1}} \left( \pi +\frac{\chi}{8} \right) \ln  \Big\lbrace \left( 16\lambda_{{1}}{r}^{2}+\pi  \right) {R}^{5}-16 \lambda_{{1}}{r}^{4}{R}^{3}-5M\pi {R}^{2}{r}^{2}+3M\pi {r}^{4} \Big\rbrace\nonumber \\
& \qquad\hspace{-1cm} +12\nu_{{2}}{\rm arctanh} \left({\frac {\lambda_{{1}}{R}^{5}-2\,\lambda_{{1}}{r}^{2}{R}^{3}-{\frac {5M\pi {R}^{2}}{16}}+\frac{3}{8}M\pi {r}^{2}}{\nu_{{1}}}  }\right)-12\nu_{{2}}\,{\rm arctanh}\left\lbrace{\frac {{R}^{2} \left( -16{R}^{3}\lambda_{{1}}+M\pi 
 \right) }{16\nu_{{1}}}}\right\rbrace \nonumber \\
& \qquad\hspace{-1cm}+8\nu_{{1}} \Big\lbrace  \left(\pi +\frac{\chi}{8}\right) \ln  \lbrace {R}^{5}\pi \left( 1-\frac{2M}{R}\right)\rbrace+\frac{3}{2}\left( \pi +\frac{\chi}{3}\right) \ln  \left( 1-{\frac {2M}{R}} \right)  \Big\rbrace  \Bigg] ,\\\label{3.3}
 & \qquad \rho^{eff}\left( r \right)={\frac {-48\,\lambda_{{1}}{R}^{5}+80\,\lambda_
{{1}}{r}^{2}{R}^{3}+15\,M\pi \,{R}^{2}-15M\pi {r}^{2}}{8{\pi }^{2}{
R}^{5}}},\\\label{3.4}
& \qquad p^{eff} \left( r \right) =-{\frac {10 \left( {R}^{3}\lambda_{{1}}-3/16M\pi 
 \right)  \left( R-r \right)  \left( R+r \right) }{{R}^{5} \left( 3 \pi +\chi \right) \pi }},
 \end{eqnarray}
 where $\lambda_{{1}}= \left( \pi +\frac{\chi}{4}\right)  \left( \pi +\frac{\chi}{2}\right)B$, $\nu_{{1}}={R}^{2}\sqrt {  {\lambda_{{1}}}^{2}{R}^{6}+\frac{1}{4}\lambda_{{1}}\pi {R}^{4}-\frac{5}{8} M\lambda_{{1}}\pi {R}^{3}-{\frac {3\,{\pi }^{2}RM}{64}}+{\frac {25{\pi }^{2}{M}^{2}}{256}}}$ and $\nu_{{2}}=\frac{4}{3}\left[ B{\pi }^{2}{R}^{3}+ \left( \frac{5}{8}B{R}^{3}\chi-{\frac {5M}{32}} \right) \pi +\frac{1}{16}B{R}^{3}{\chi}^{2} \right] 
 \left(\pi +\frac{\chi}{4}\right) {R}^{2}$.


\begin{figure}[!htp]\centering
	\includegraphics[scale=.3]{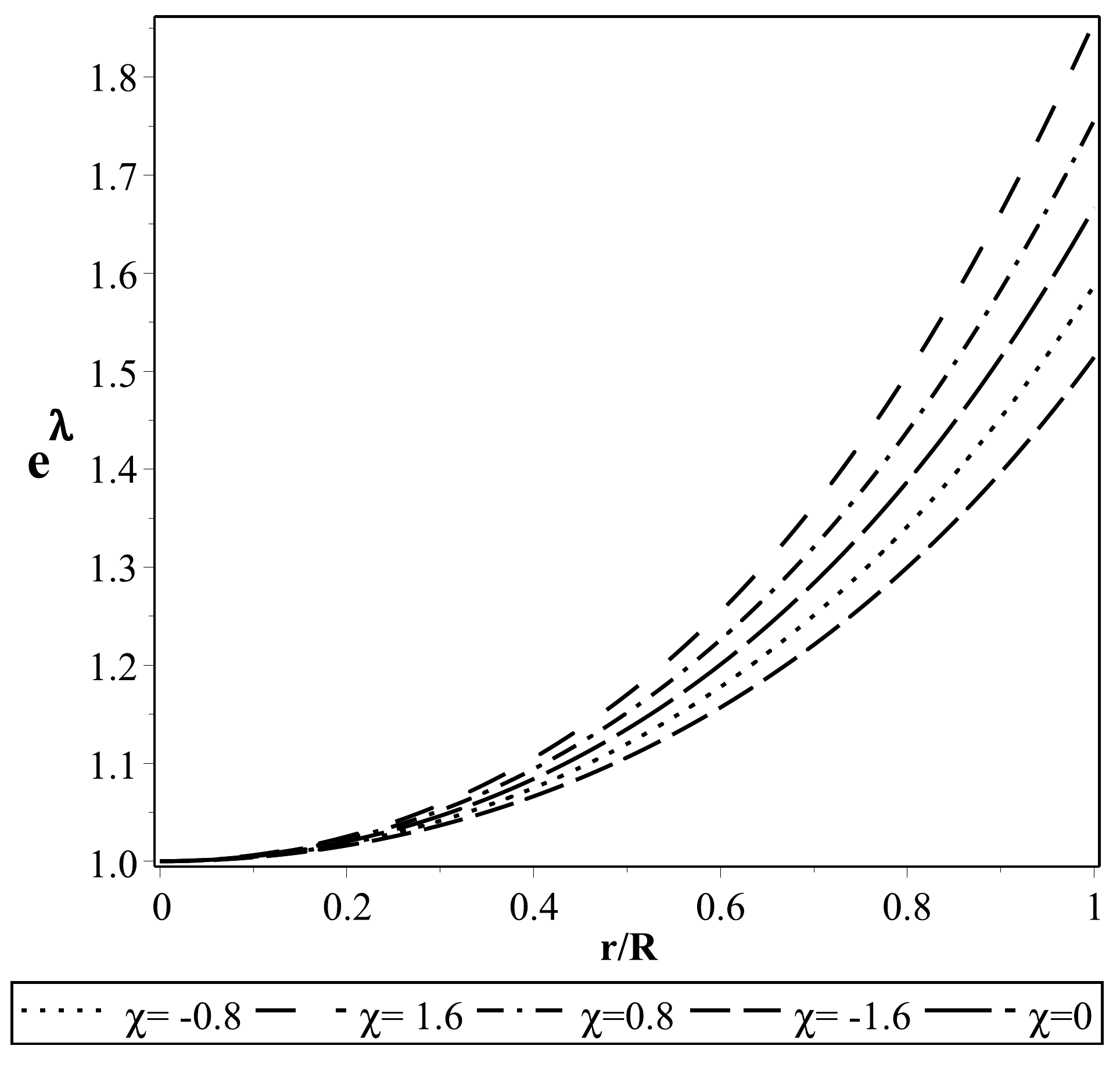}
	\includegraphics[scale=.3]{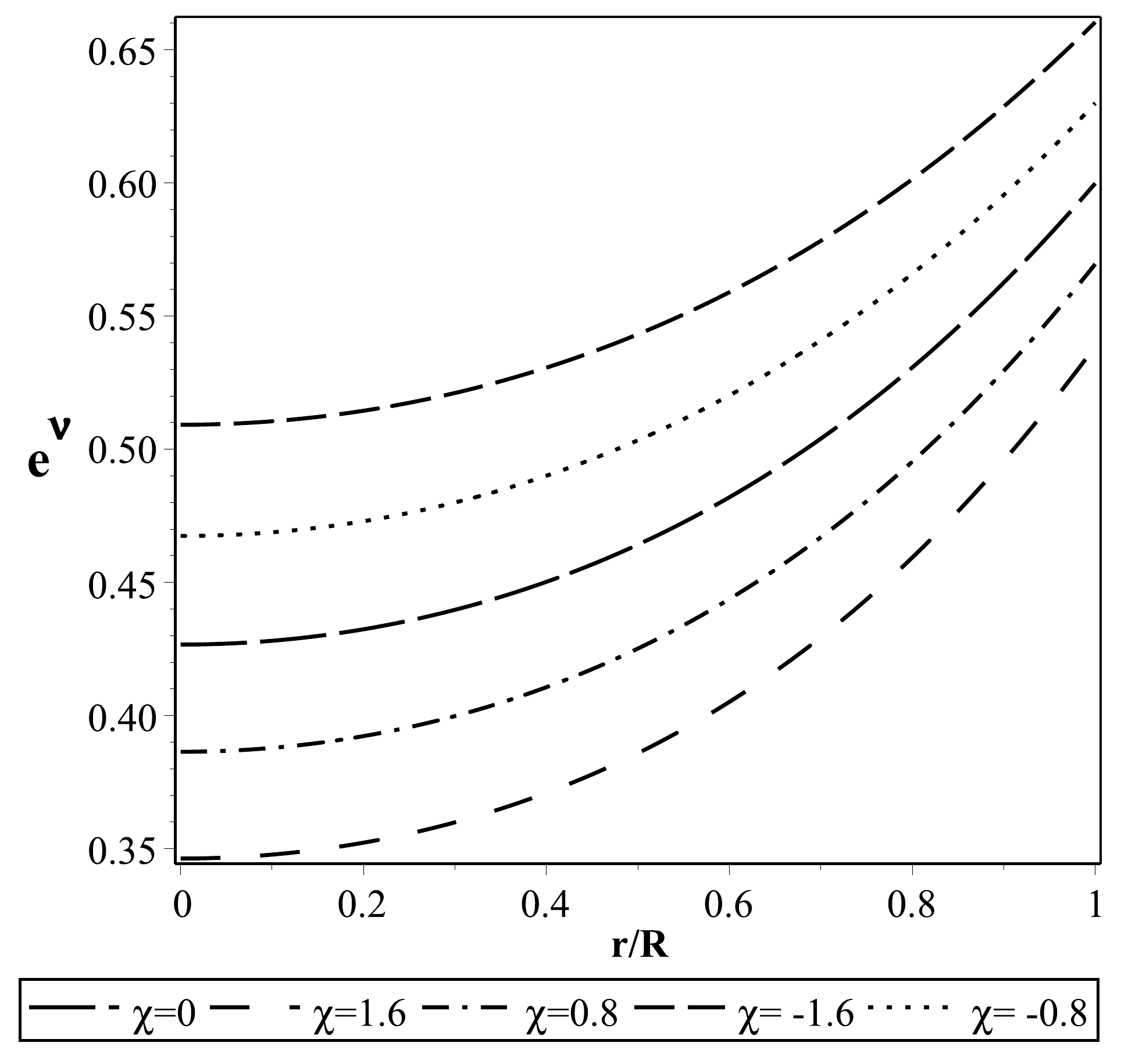}
	\caption{Variation of (i) ${e}^{\lambda(r)}$ (left panel) and (ii) ${e}^{\nu(r)}$
(right panel) as a function of the fractional radial coordinate $r/R$ for the strange star candidate $LMC~X-4$ are shown. Here and in what follows $B=83~ MeV/{{fm}^3}$.} \label{Fig1}
\end{figure}

 

\begin{figure}[!htp]\centering
\includegraphics[scale=.3]{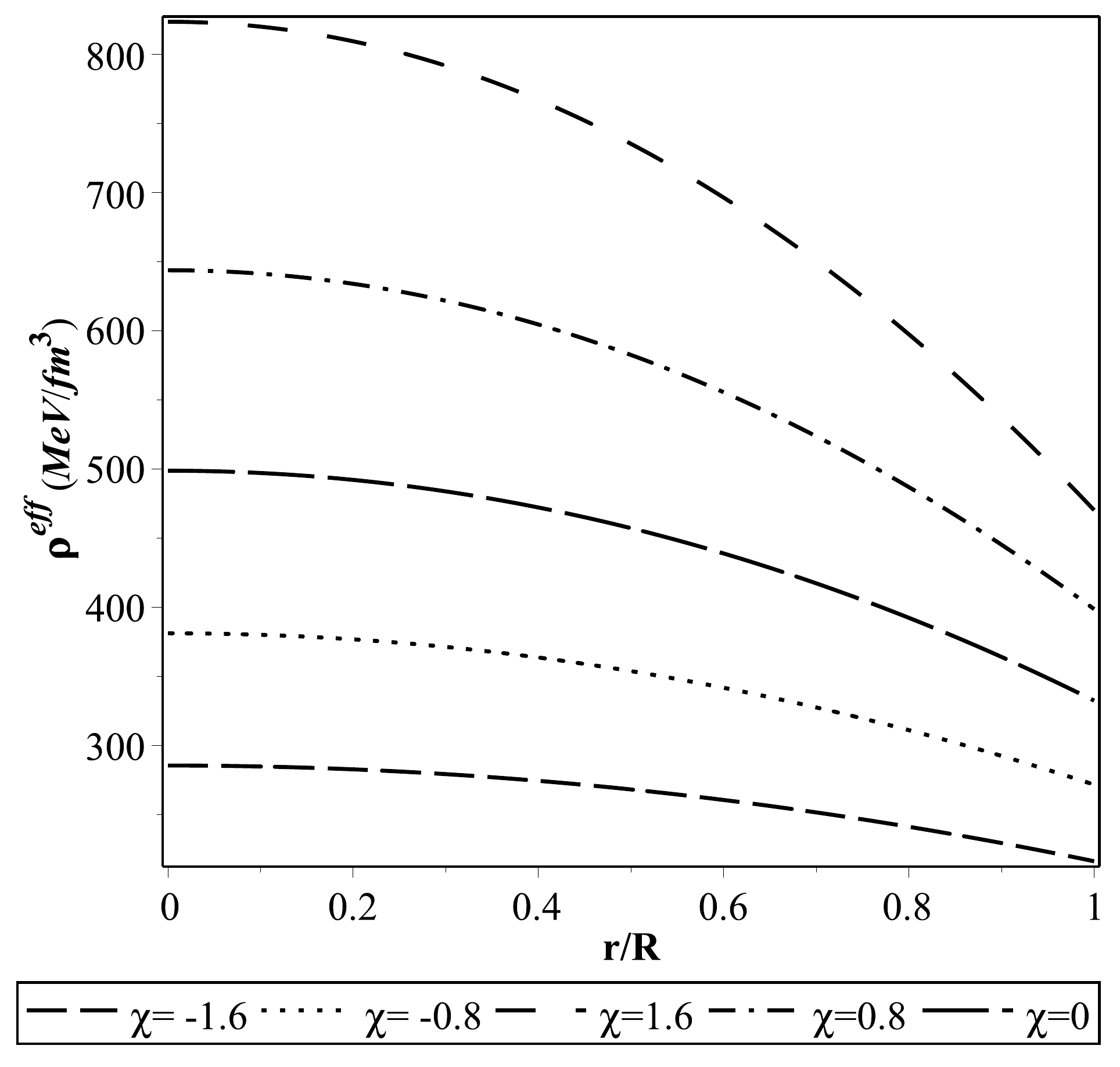}
\includegraphics[scale=.3]{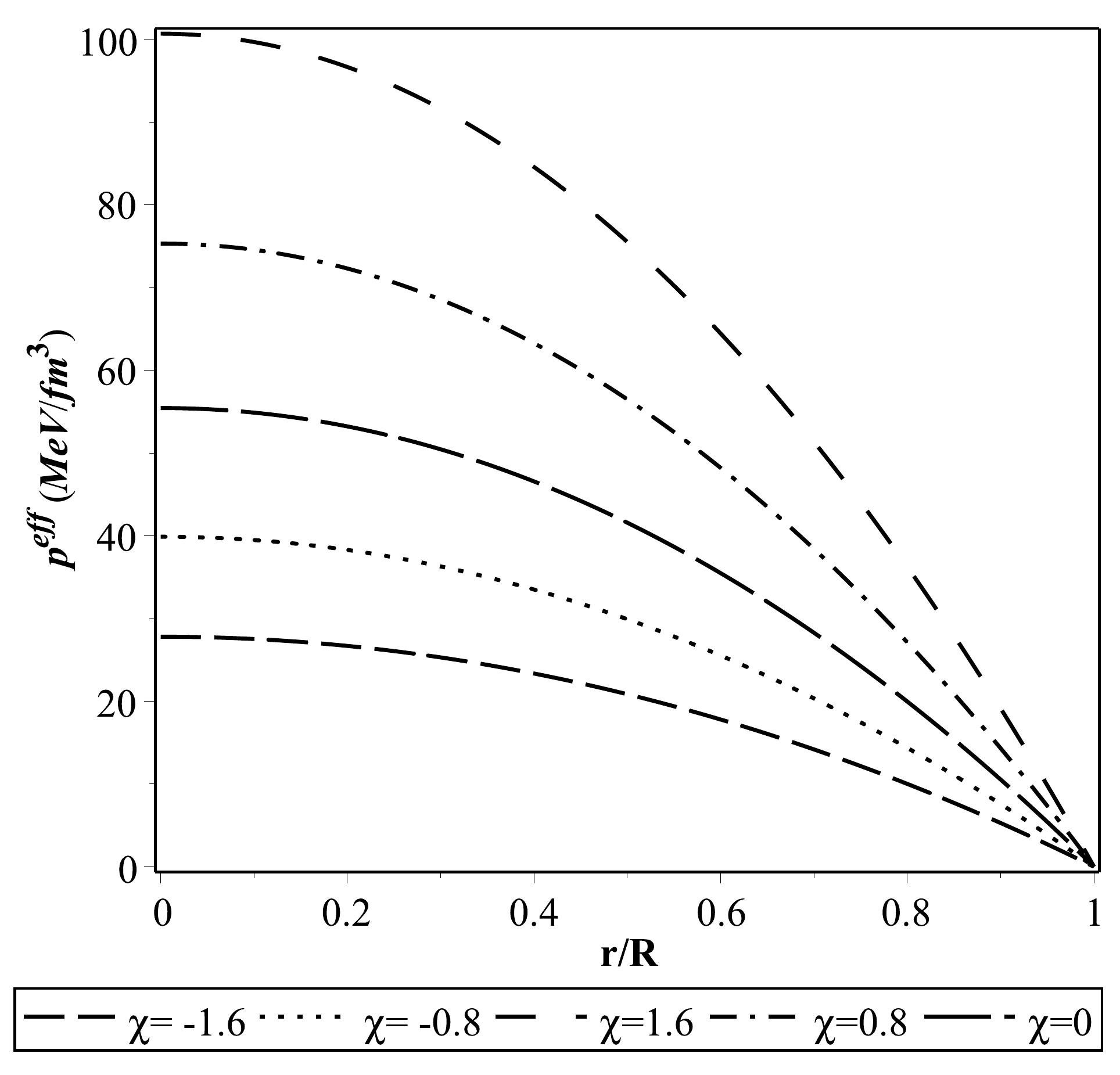}
\caption{Variation of (i) the density $\mathbf{\rho^{eff}}$ (left panel) and (ii) pressure $\mathbf{p^{eff}}$ (right panel) as a function of the fractional radial coordinate $r/R$ for the strange star candidate $LMC~X-4$ are shown.}  \label{Fig2}
\end{figure}

 
The variation of the physical parameters, viz., $e^{\lambda}$, $e^{\nu}$, $\mathbf{{\rho}^{eff}}$ and $\mathbf{p^{eff}}$ are featured in figures~\ref{Fig1} and \ref{Fig2}. 
 
We find from eq. (\ref{1.6}) the modified form of equation of conservation due to $f(R,\mathcal{T})$ gravity as follows
 \begin{eqnarray}\label{3.5}
 -p^{{\prime}}-\frac{1}{2}\nu^{{\prime}} \left( \rho+p \right) +{\frac {\chi}{8\,\pi +2\,\chi}}\left( 
\rho^{{\prime}}-p^{{\prime}} \right)=0.
\end{eqnarray}  
 
 Using eqs. (\ref{2.3}), (\ref{2.7}), (\ref{2.11}) and (\ref{3.5}) we have hydrostatic equilibrium equation for $f\left(R,\mathcal{T}\right)$ gravity which can be given by
  \begin{eqnarray}\label{3.6}
 p^{{\prime}}=-{\frac{(\rho+p) \left[ 4\pi pr+{\frac {m}{{r}^{2}}}-\frac{1}{2}\chi\,
 \left( \rho-3p \right) r \right]}{\left( 1-{\frac {2m}{r}}
 \right) \left[ 1+{\frac {\chi}{8\pi +2\chi}}\left( 1-{\frac {\rm d\rho}{{\rm d}
p}}\right) \right]}},
\end{eqnarray}  
 where we have considered that the matter-energy density $\rho$ depends on the pressure $p$, i.e., $\rho=\rho\left(p\right)$. 

By substituting $\chi=0$ in eq.~(\ref{3.6}) the standard form of the Tolman-Oppenheimer-Volkoff (TOV) equation can be achieved  as applicable in GR. Now using MIT bag EOS~(\ref{2.6}) and eq.~(\ref{2.8}) we have presented exact solution for the hydrostatic equilibrium equation~(\ref{3.6}) for $f\left(R,\mathcal{T}\right)$ gravity theory. Here, as mentioned earlier, the value of the bag constant is assumed as $B=83~MeV/{{fm}^3}$~\cite{Rahaman2014}.  We have predicted radius of the strange star candidate $LMC~X-4$ for the parametric values of $\chi$ in Table~\ref{Table 1} and values of the observed mass as shown in Table~\ref{Table 2}. 


{\footnotesize{\begin{table}[htp]
  \centering
    \caption{Numerical values of the physical parameters for the strange star candidate $LMC~X-4$ having mass~$1.29 \pm 0.05~{M_{\odot}}$~\cite{dey2013} due to the different values of $\chi$.} \label{Table 1}
    \resizebox{\columnwidth}{!}{
    \begin{tabular}{ *6{>{\centering\arraybackslash}m{1in}} @{}m{0pt}@{}}
    \hline
Values of $\chi$ & $\chi=-1.6$ & $\chi=-0.8$ & $\chi=0$ & $\chi=0.8$ & $\chi=1.6$ &\\ [2ex]
\hline
Predicted Radius \textit{(Km)} & 11.205 & 10.285 & 9.511 & 8.841 & 8.249 &\\[1ex]

${\rho}^{eff}_c$ ($gm/{{cm}^3}$) & $5.090\times {{10}^{14}}$ & $6.795\times {{10}^{14}}$ & $8.892\times {{10}^{14}}$ & $1.148\times {{10}^{15}}$ & $1.468\times {{10}^{15}}$ &\\ [1ex]

${\rho}^{eff}_0$ ($gm/{{cm}^3}$) & $3.855\times {{10}^{14}}$ & $4.843\times {{10}^{14}}$ & $5.927\times {{10}^{14}}$ & $7.107\times {{10}^{14}}$ & $8.383\times {{10}^{14}}$ &\\ [1ex]

${p}^{eff}_c$~($dyne/{{cm}^2}$) & $4.455\times {{10}^{34}}$ & $6.391\times {{10}^{34}}$ & $8.883\times {{10}^{34}}$ & $1.207\times {{10}^{35}}$ & $1.613\times {{10}^{35}}$ &\\ [1ex]

${2M}/{R}$ & 0.34 &  0.37 & 0.40 & 0.43 & 0.46 &\\[1ex]

$Z_s$ & 0.23 & 0.26 & 0.29 & 0.33 & 0.36 &\\[1ex]
\hline
  \end{tabular}
  }
  \end{table}
}}



\begin{table}[htp]
  \centering
    \caption{Numerical values of physical parameters for the different strange star candidates for $\chi=-0.8$ } \label{Table 2}
    \resizebox{\columnwidth}{!}{
    \begin{tabular}{ *7{>{\centering\arraybackslash}m{1in}} @{}m{0pt}@{}}
    \hline
Strange Stars & Observed Mass ($M_{\odot}$) & Predicted Radius ($Km$)
 & ${{\rho}^{eff}_c}$ $(gm/{cm}^3)$ & ${{p}^{eff}_c}$ $(dyne/{cm}^2)$ & Surface Redshift & $\frac{2M}{R}$ \\ 
\hline 
$Vela~X-1$ & $1.77 \pm 0.08$~\cite{dey2013} & $11.180 \pm 0.122$ & $7.756\times {{10}^{14}}$ & $9.534\times {{10}^{34}}$ & 0.370 & 0.467 \\ 
 
$4U~1820-30$ & $1.58 \pm 0.06$~\cite{guver2010b} & $10.866 \pm 0.107$ & $7.339\times {{10}^{14}}$ & $8.172\times {{10}^{34}}$ & 0.323 & 0.429 \\ 
  
$Cen~X-3$ & $1.49 \pm 0.08$~\cite{dey2013} & $10.700 \pm 0.153$ & $7.159\times {{10}^{14}}$ & $7.580\times {{10}^{34}}$ & 0.303 & 0.411 \\ 
 
$LMC~X - 4$ & $1.29 \pm 0.05$~\cite{dey2013} & $10.285 \pm 0.112$ & $6.796\times {{10}^{14}}$ & $6.393\times {{10}^{34}}$ & 0.260 & 0.370 \\ 
 
$SMC~X - 1$ & $1.04 \pm 0.09$~\cite{dey2013} & $9.667 \pm 0.247$ & $6.387\times {{10}^{14}}$ & $5.054\times {{10}^{34}}$ & 0.210 & 0.317 \\ 
\hline
  \end{tabular}
  }
  \end{table}


\section{Physical properties of the stellar model under $f\left(R,\mathcal{T}\right)$ gravity}\label{sec4}
In this section we discuss physical validity of the achieved solutions under $f\left(R,\mathcal{T}\right)$ gravity. By studying the energy conditions, modified form of the TOV equation, the adiabatic index etc., we try to explore different physical properties of the stellar model.
 
 \subsection{Energy conditions}\label{subsec4.1}
 To be consistent with the energy conditions, viz., Null Energy Condition (NEC), Weak Energy Condition (WEC), Strong Energy
Condition (SEC) and Dominant Energy Condition (DEC) in $f\left(R,\mathcal{T}\right)$ gravity theory, following inequalities should be hold simultaneously for the system given as~\cite{SC2013}
 {{\begin{eqnarray}\label{4.1.1}
&\qquad\hspace{-1.5cm} ~NEC:~~{{\rho}^{eff}}+{{p}^{eff}}\geq 0,\\ \label{4.1.2}
&\qquad\hspace{-1.5cm} ~WEC:~~{{\rho}^{eff}}+{{p}^{eff}}\geq 0,~{{\rho}^{eff}}\geq 0, \\ \label{4.1.3}
&\qquad\hspace{-1.5cm} ~SEC:~~{{\rho}^{eff}}+{{p}^{eff}}\geq 0,~{{\rho}^{eff}}+3{{p}^{eff}}\geq 0, \\ \label{4.1.4}
&\qquad\hspace{-1cm}  ~DEC:~~{{\rho}^{eff}}\geq 0,~{{\rho}^{eff}}-{{p}^{eff}}\geq 0.
 \end{eqnarray}}}
  
\begin{figure}[!htpb]
\centering
   \includegraphics[width=5cm]{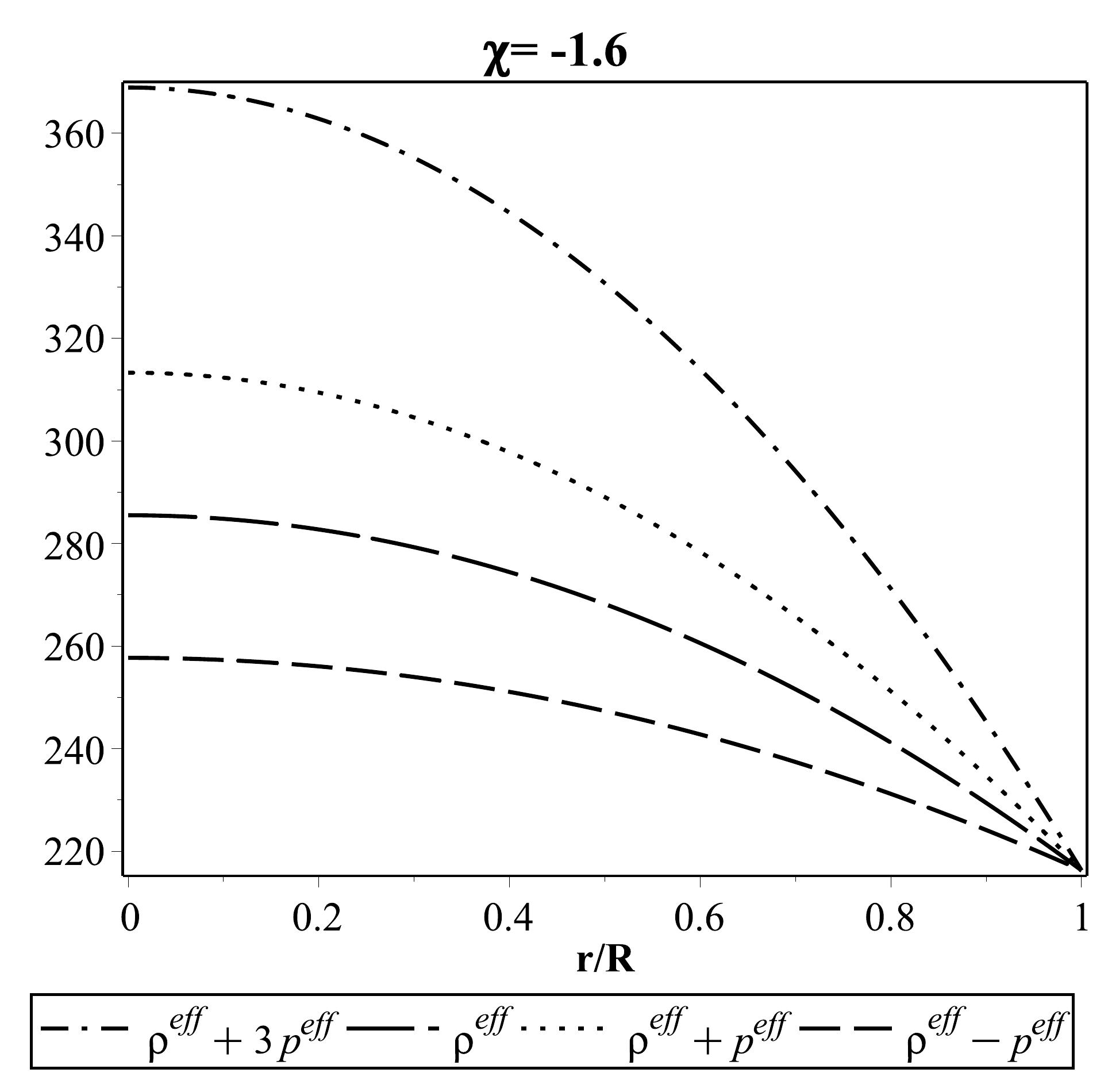}
   \includegraphics[width=5cm]{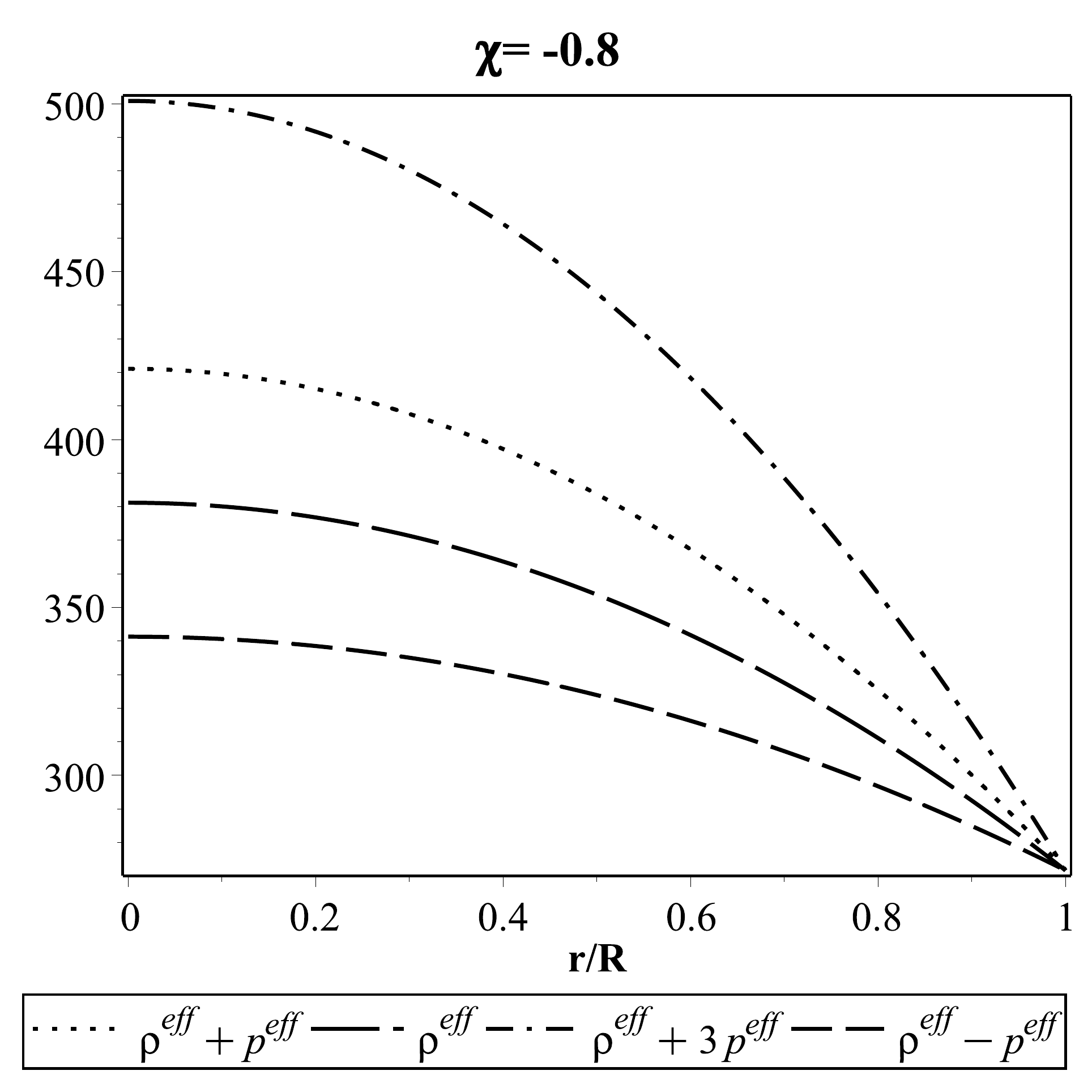}
   \includegraphics[width=5cm]{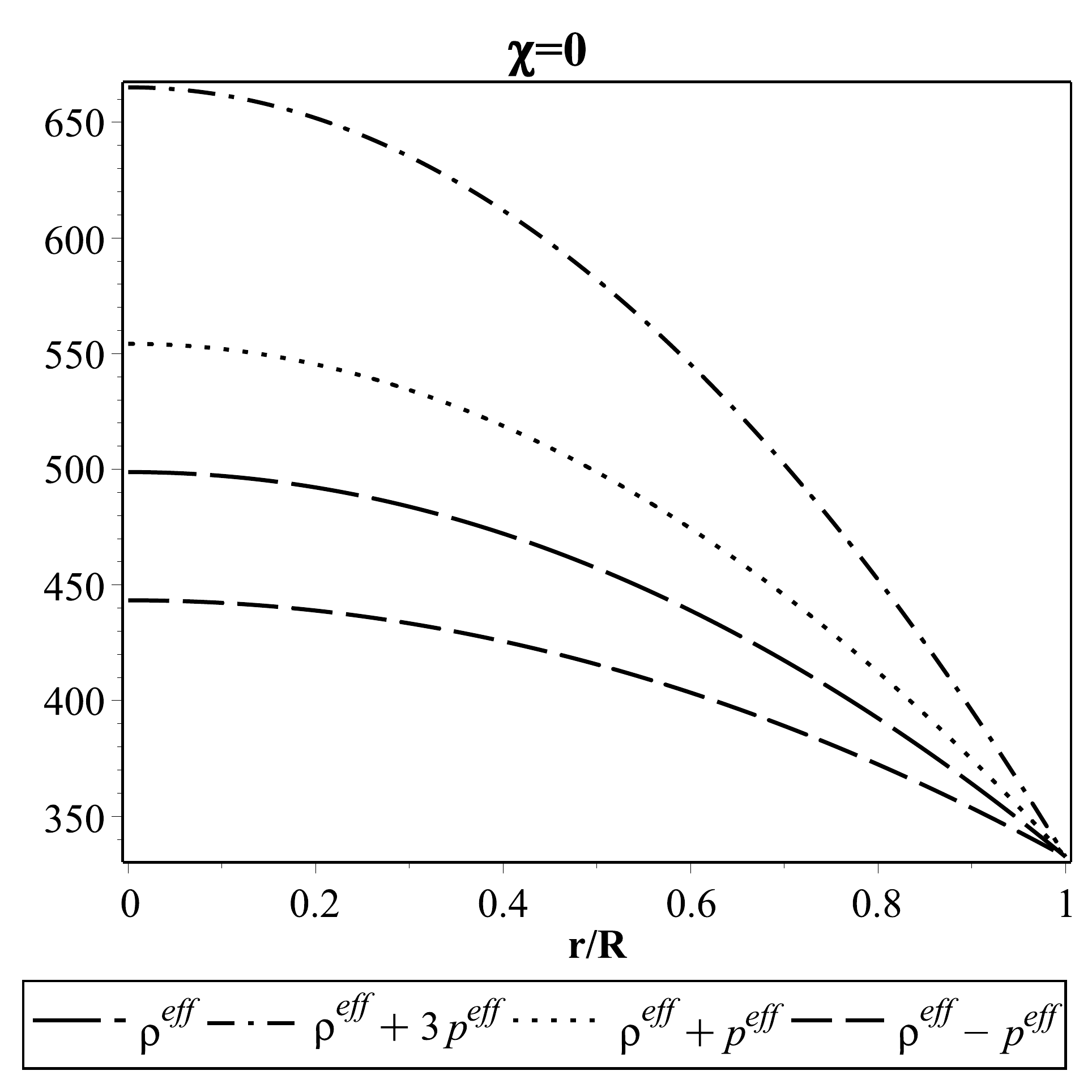}
   \includegraphics[width=5cm]{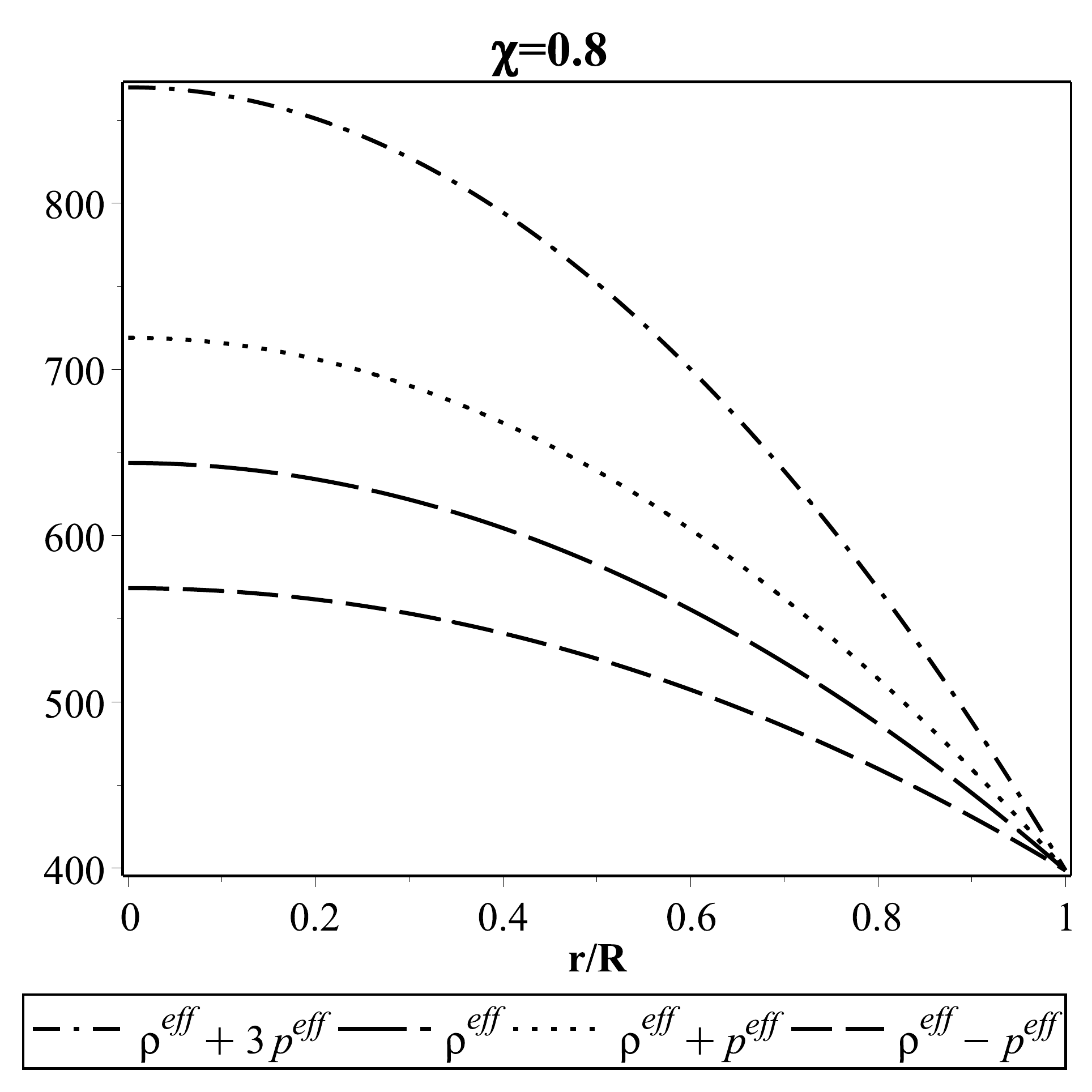}
   \includegraphics[width=5cm]{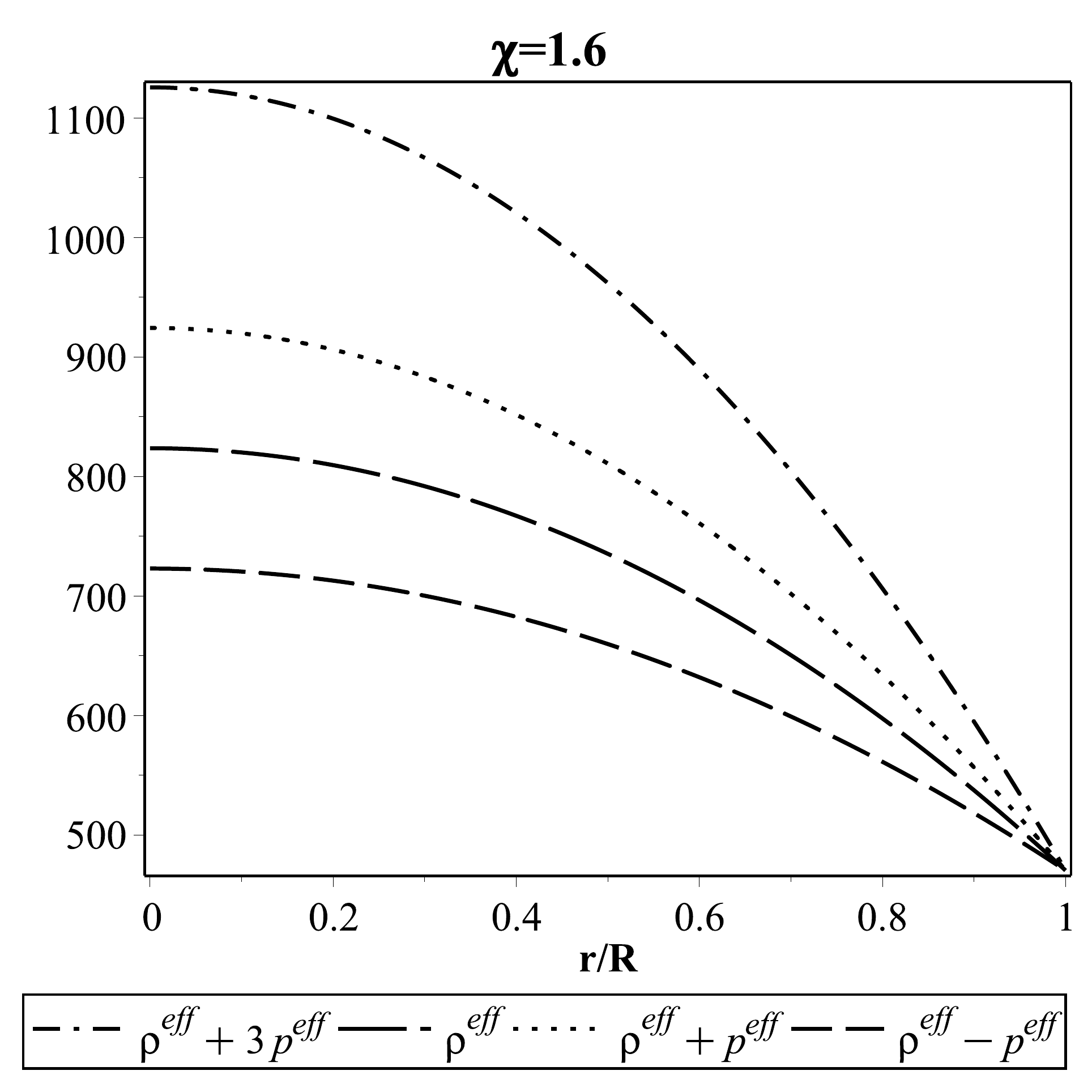}       
    \caption{Variation of the energy conditions with the radial coordinate $r/R$ for the strange star candidate $LMC~X-4$ due to different chosen values of $\chi$. } \label{Fig3}
\end{figure}

Figure~\ref{Fig3} shows that our model is consistent with all the energy conditions for all the chosen parametric values of $\chi$ and hence confirms that the physical validity of the solution.

\subsection{Mass-Radius relation}\label{subsec4.2}
Using eqs.~(\ref{2.2a}),~(\ref{2.6})~and~(\ref{2.7}) we obtain mass function for our system as 
\begin{eqnarray}\label{4.2.1}
m \left( r \right) = \left({1+{\frac {  \chi  
}{3\pi }}}\right){{\tilde{m}}}+\frac{2}{9}\chi\,B{r}^{3},
\end{eqnarray}
where $\tilde{m}=4\pi \int_{0}^{r}\!{{\rho}} \left( r \right) {r}^{2}{dr}$ representing the usual mass term due to the quark matter, whereas the extra term ${\frac {\chi}{3\pi }}\left( \tilde{m}+\frac{2}{3}\pi B{r}^{3} \right)$ is due to the effect of the modification of GR. 

Hence the total mass of the system as given in eq.~(\ref{2.10}) becomes 
\begin{eqnarray}\label{4.2.1b}
M=\frac{2R}{45\pi}}{\left(72B{\pi}^{2}{R}^{2}
+56B\pi {R}^{2}\chi+9B{R}^{2}{\chi}^{2}+12{\pi }^{2}{R}^{2}\rho_{{c}}+4\pi {R}^{2}\chi\rho_{{c}} \right).
\end{eqnarray}

\begin{figure}[!htpb]\centering
\includegraphics[width=8cm]{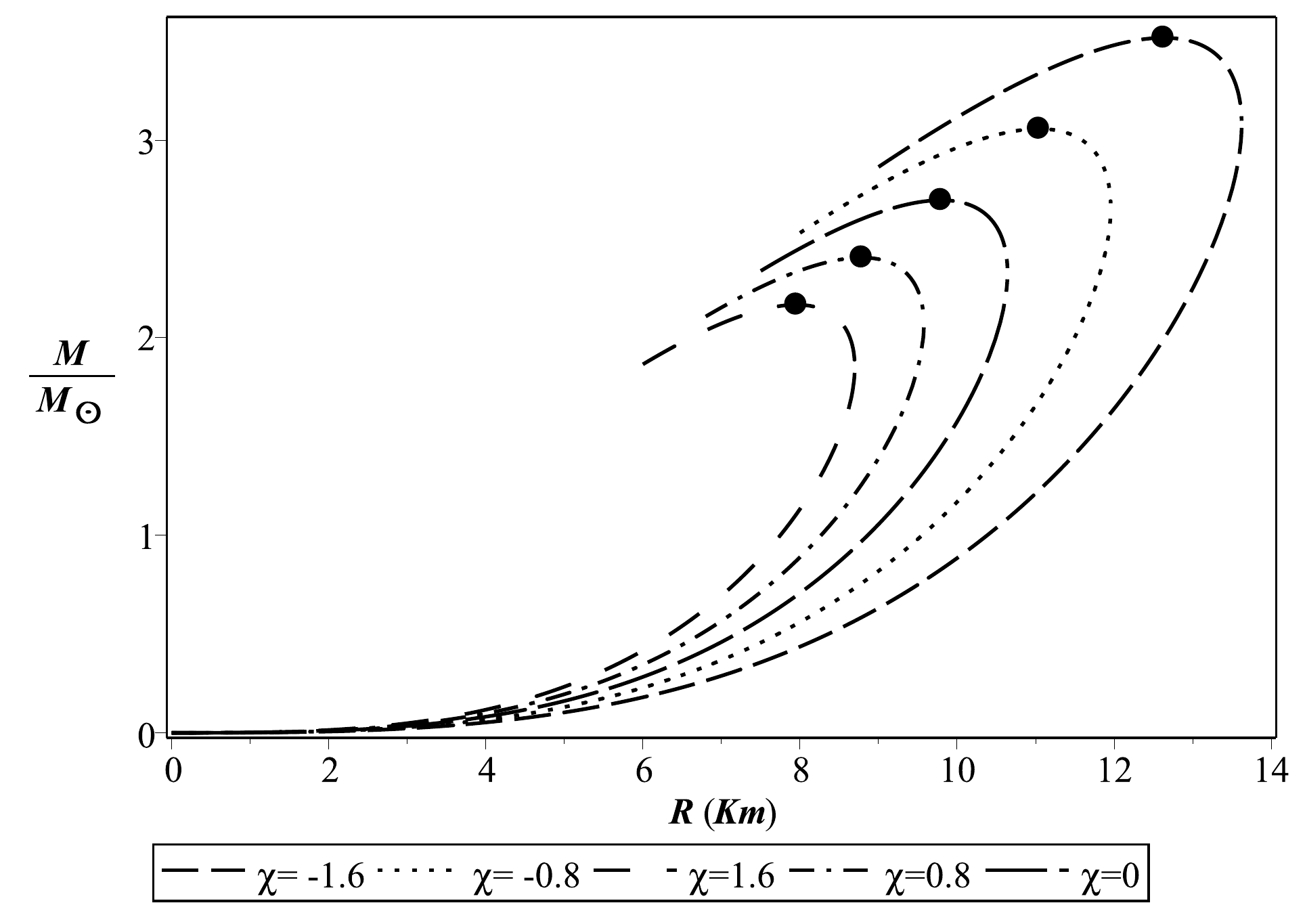}
	\caption{Mass~$(M/{M_{\odot}})$ vs Radius~($R$~in km) curve of the strange star candidates for different values of $\chi$. Solid circles are representing maximum mass for the strange star candidates.}  \label{Fig4}
\end{figure}

In figure~\ref{Fig4}, for the chosen values of $\chi$, viz. $-1.6,~-0.8,~0,~0.8$ and $1.6$, we have presented the total mass~$M$ (normalized in solar mass) versus the radius~$R$ relation for the strange star candidates. This figure~\ref{Fig4} features that the maximum value of mass~$\left(M_{max}\right)$ gradually decreases for the increasing values of $\chi>0$. On the other hand, $M_{max}$ increases with the decreasing values of $\chi<0$.


\begin{figure}[!htpb]
\centering
\includegraphics[width=6cm]{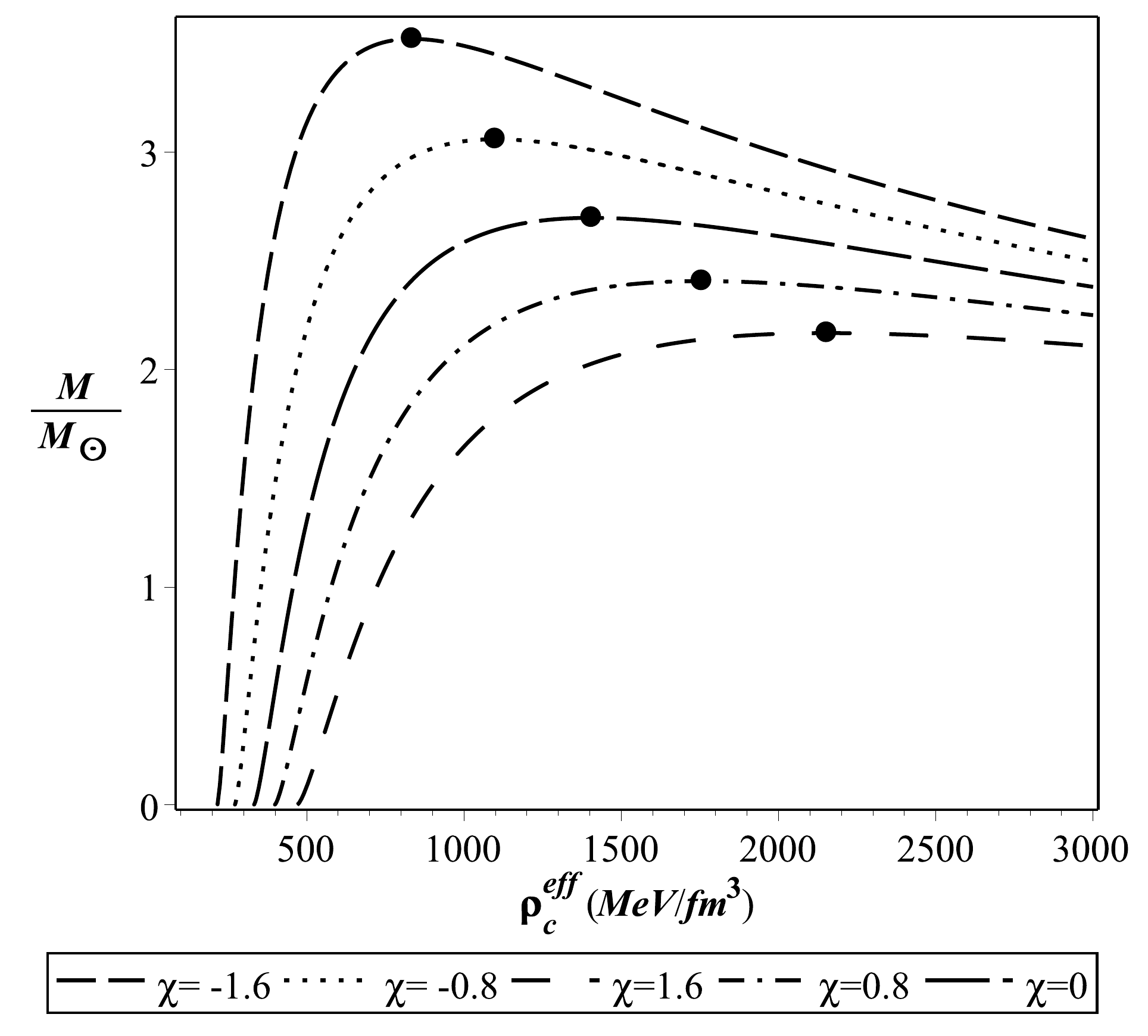}
\includegraphics[width=6cm]{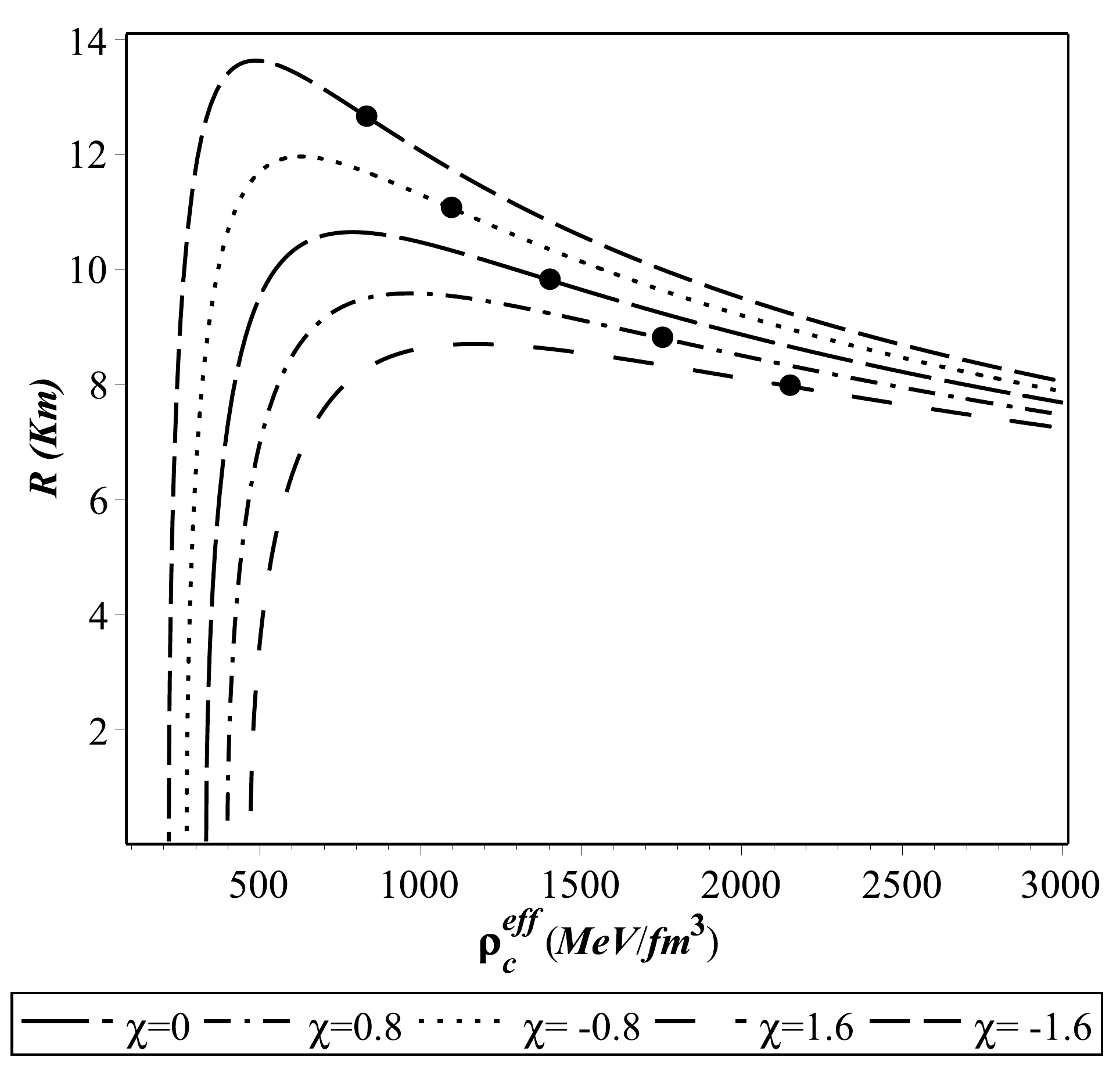}
\caption{Variation of the (i) mass~$M/{M_{\odot}}$ (left panel) and (ii) radius~$R$~in km (right panel) of the strange star candidates as a function of the central density $({\rho^{eff}_c})$ are shown. Solid circles are representing maximum mass for the strange star candidates.}  \label{Fig5}
\end{figure}


In figure~\ref{Fig5} we have shown variation of mass~($M/{M_{\odot}}$ in the left panel) and radius~($R$ in the right panel) for the strange star candidates with respect to the central density~$\left({\rho}^{eff}_c\right)$. We find that as the value of $\chi$ increases gradually from the negative to the positive values, the maximum mass points are achieved for the higher values of ${\rho}^{eff}_c$. Also, we find that due to decreasing values of $\chi$ from $\chi=0$, the value of the maximum mass gradually increases and with the increasing values of $\chi$ from $\chi=0$, the value of the maximum mass gradually decreases. For example, when $\chi=-1.6$ and $1.6$ the maximum masses~$3.52~{M_{\odot}}$ and ~$2.17~{M_{\odot}}$ are achieved for ${\rho}^{eff}_c=4.774\times{{10}^{14}}$~$gm/{{cm}^3}$ and ${\rho}^{eff}_c=1.230\times{{10}^{15}}$~$gm/{{cm}^3}$, respectively. When $\chi=0$ the maximum mass~$2.70~{M_{\odot}}$ is attained for ${\rho}^{eff}_c=2.508\times{{10}^{15}}$~$gm/{{cm}^3}$. Figure~\ref{Fig5}~(right panel) also features that the total radius $R$ gradually decreases with the increase of $\chi$ from the negative to the positive values. When $\chi=-1.6$ and $1.6$, the radius corresponding to the maximum mass points are given as $12.630~km$ and $7.955~km$, respectively. On the other hand, for $\chi=0$ we find the radius corresponding to the maximum mass point is $9.797~km$. So, as $\chi$ gradually increases from $\chi=0$ the strange star candidates become less massive and smaller but they become more compact due to increase of the density, however when $\chi$ decreases from $\chi=0$ the strange star candidates become more massive, larger in radius but less dense stellar system. Thus, we are able to justify the main motivation for working with the alternative gravity theories in the context of stellar hydrostatic equilibrium that our model allows more massive objects (in comparison to GR) to exist.

\subsection{Compactification factor and redshift}\label{subsec4.3}
The compactification factor for our model in $f\left(R,\mathcal{T}\right)$ gravity is given as 
\begin{eqnarray}\label{4.3.1}
u\left(r\right)=\frac{m\left(r\right)}{r}={\frac {{r}^{2} \left( -16\lambda_{{1}}{R}^{5}+16\lambda_{{1}
}{r}^{2}{R}^{3}+5M\pi {R}^{2}-3M\pi {r}^{2} \right) }{2\pi{R}^{5}}}.
\end{eqnarray}


\begin{figure}[!htpb]\centering
\includegraphics[scale=.3]{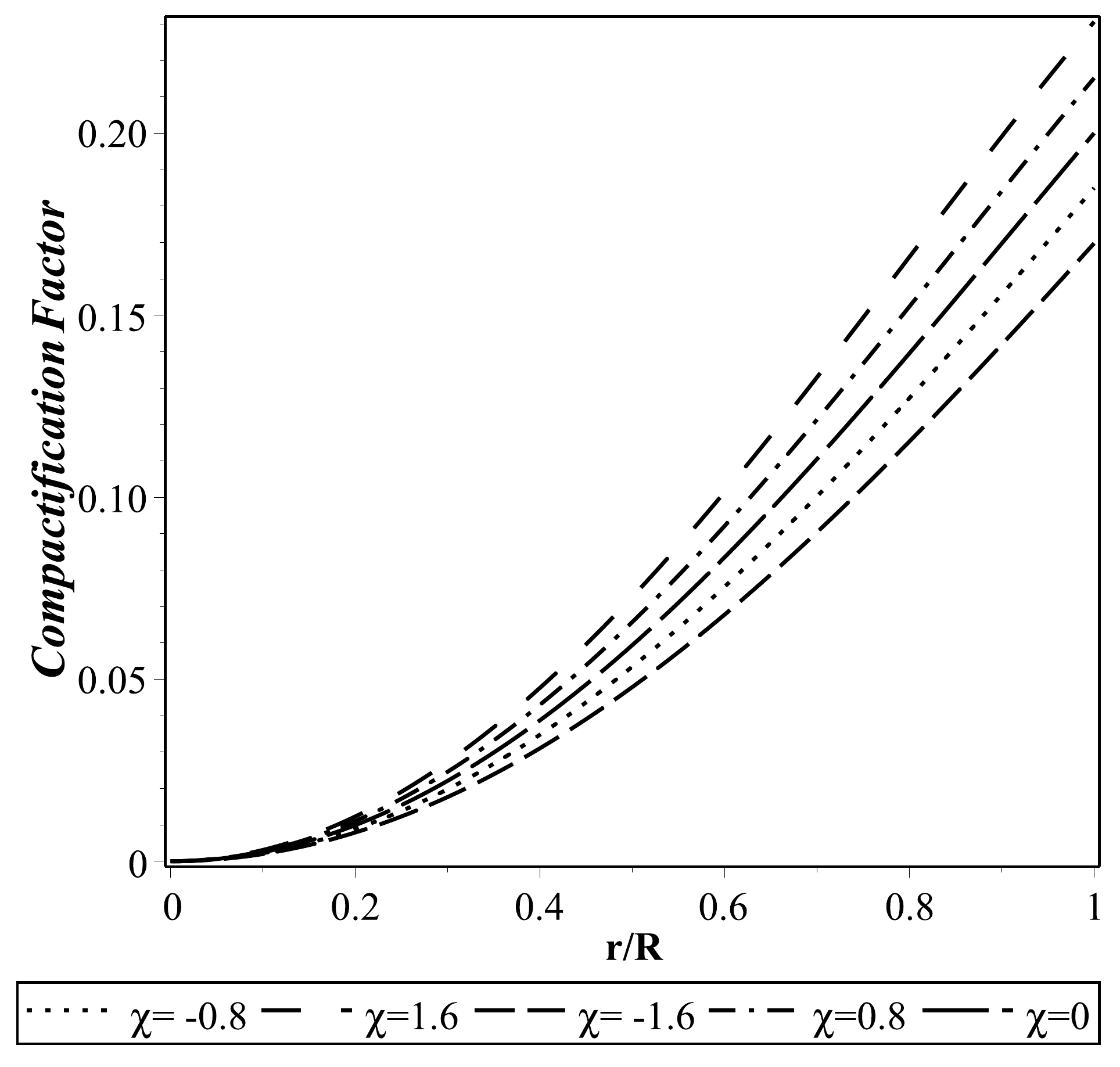}
\includegraphics[scale=.3]{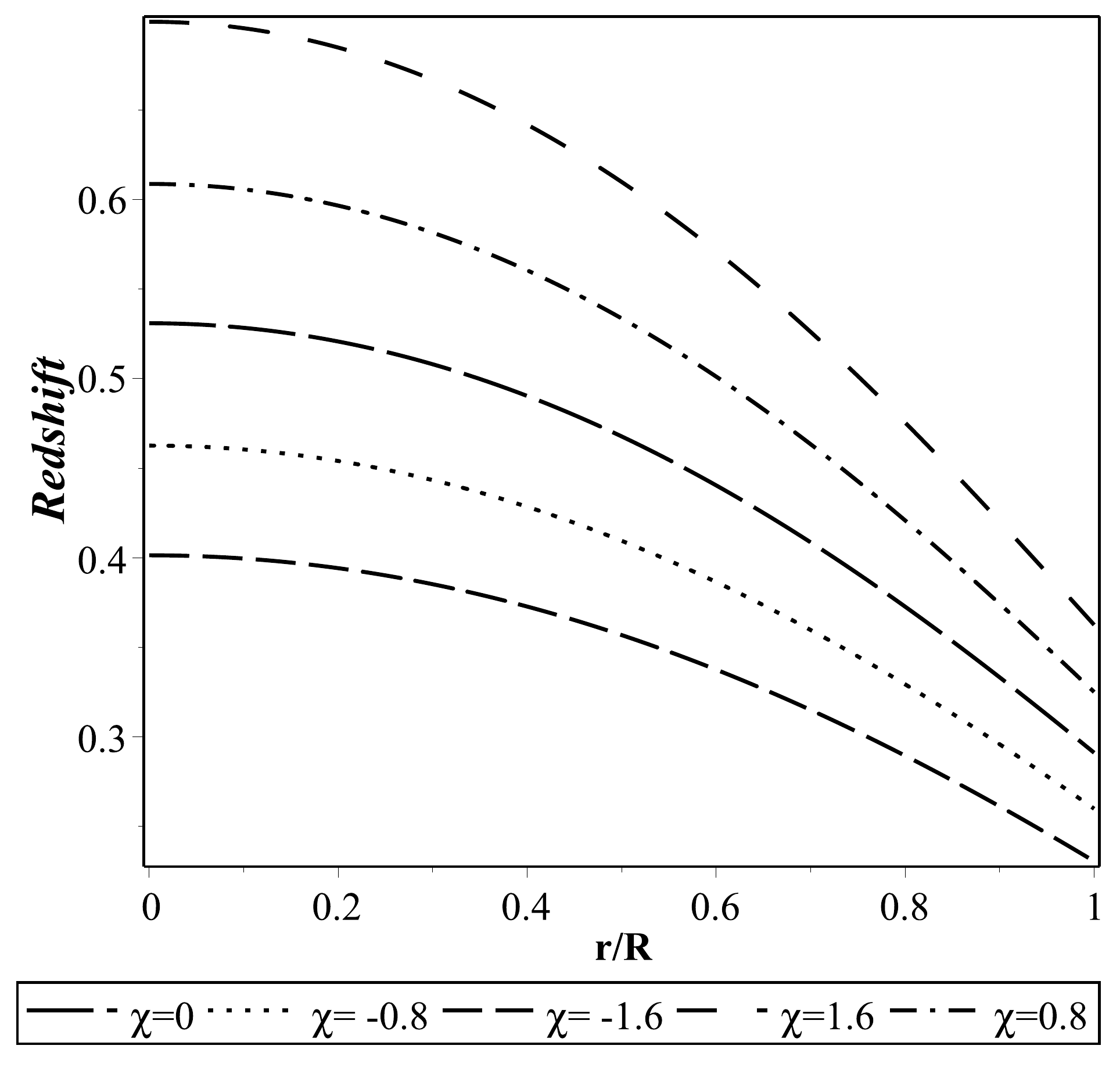}
\caption{Variation of the (i) compactification factor (left panel) and (ii) redshift (right panel) as a function of the radial coordinate $r/R$ for the strange star candidate $LMC~X-4$.}  \label{Fig6}
\end{figure}


The surface redshift is defined as
\begin{eqnarray}
& \qquad {Z_s}={{\rm e}^{-\frac{\nu \left( R \right)}{2} }}-1={\frac {1}{\sqrt {1-{\frac {288B{\pi }^{2}{R}^{2}+224B\pi {R}^{2}\chi+36B{R}^{2}{\chi}^{2}+48{\pi }^{2}{R}^{2}\rho_{{c}}+16\pi 
{R}^{2}\chi \rho_{{c}}}{45\pi }}}}}-1.
\end{eqnarray}

It is known that one can quantify the gravitational redshift of an absorption line that only allows the measurement of the compactness, but not mass and radius distinctly unless other advanced methods are used. However, in the present study we find from figure~\ref{Fig6} that as the value of $\chi$ decreases from $\chi=0$, the compactification factor as well as the redshift of the stellar system decrease consequently. On the other hand, when $\chi$ increases from $\chi=0$ we find that both the factors increase accordingly. The profiles of both the compacification factor and redshift function are featured in figure~\ref{Fig6}.

\subsection{Stability of the system}\label{subsec4.4}
In the following subsections we have discussed stability of the stellar system under $f\left(R,\mathcal{T}\right)$ gravity theory by studying (i)~Modified TOV equation, (ii)~Causality condition and (iii)~Adiabatic index. 

\subsubsection{Modified TOV equation}\label{subsubsec4.4.1}
We have shown modified form of the energy conservation equation for $f\left(R,\mathcal{T}\right)$ gravity in eq.~(\ref{1.6}) and later in a more concise form in eq.~(\ref{1.7}). This clearly invites modification of the standard form of the Tolman-Oppenheimer-Volkoff (TOV) equation~\cite{Varela2010,Ponce1993} for GR. In the present case modified form of the TOV equation is shown in eq.~(\ref{3.5}), given as
\begin{equation*}
-p^{{\prime}}-\frac{1}{2}\nu^{{\prime}} \left( \rho+p \right) +{\frac {\chi}{8\pi +2\chi}}\left( 
\rho^{{\prime}}-p^{{\prime}} \right)=0,
\end{equation*}
where the first term represents the hydrodynamic force $(F_h)$, second term is gravitational force~$(F_g)$ and the last term represents a force~$(F_m)$ which arises due to modification of the gravitational Lagrangian of the Einstein-Hilbert action. 

Above eq.~(\ref{3.5}) predicts that for the system to be in equilibrium the sum of different forces for our system must be zero, i.e., ${F_h}+{F_g}+{F_m}=0$. For our system the forces are as follows:
\begin{eqnarray}\label{4.4.1.1}
&\qquad\hspace{-1cm} {F_{{g}}=-\frac{1}{2}{{\nu}^{\prime}}\left(\rho+p\right)=\big[10 \big\lbrace  \left[ B{R}^{3}{\pi }^{2}+ \left( \frac{3}{4}B{R}^{3}\chi-\frac{3}{16}M \right) \pi +\frac{1}{8}B{R}^{3}{\chi}^{2} \right] {r}^{2}-{\frac {7f_{{1}}}{10}} \big\rbrace} \nonumber\\ 
&\qquad {\big\lbrace  \left( \frac{\chi}{8}+\pi  \right)  \left( -\frac{16}{3}
{R}^{3}\lambda_{{1}}+M\pi  \right) {r}^{2}+\frac{16}{3}f_{{2}} \big\rbrace r
\big]{\Big/}\big[3 \left( \pi +\frac{\chi}{3}\right) ^{2}\pi {R}^{5} \big\lbrace  \left( -{R}^{3}\lambda_{{1}}+\frac{3}{16}M\pi  \right) {r}^{4}}\nonumber\\
& \qquad {+\left[ B{R}^{3}{\pi }^{2}+ \left( \frac{3}{4}B{R}^{3}\chi-{\frac {5M}{16}} \right) \pi +\frac{1}{8}B{R}^{3}{\chi}^{2} \right]{R}^{2}{r}^{2}+\frac{1}{16}\,{R}^{5}\pi  \big\rbrace\big]},\\\label{4.4.1.2}
& \qquad\hspace{-5.5cm} F_{{h}}=-\frac{dp}{dr}=-{\frac {5r \left( 16B{R}^{3}{\pi }^{2}+12B\pi {R}^{
3}\chi+2B{R}^{3}{\chi}^{2}-3M\pi  \right) }{4\pi{R}^{5} \left( 3
\pi +\chi \right) }},\\\label{4.4.1.3}
&\qquad\hspace{-4.8cm} F_{{m}}={\frac {\chi}{8\pi +2\chi}}\left(\rho^{{\prime}}-p^{{\prime}} \right)={\frac {5\chi r \left( 16B{R}^{3}{\pi }^{2}+12B\pi {R}^{3}\chi+2B{R}^{3}{\chi}^{2}-3M\pi  \right) }{ 4\left( 4\pi +
\chi \right) \pi {R}^{5} \left( 3\pi +\chi \right) }},
\end{eqnarray}
where $f_{{1}}= \left( \frac{\chi}{2}+\pi  \right)  \left( \pi +\frac{3\chi}{14} \right) B{R}^{5}-{\frac {15}{56}}M\pi {R}^{2}$ and $f_{{2}}= \lambda_{{1}}\left( \frac{\chi}{8}+\pi  \right) {R}^{5}-{\frac {15}{64}}M \left( \pi +\frac{\chi}{6} \right) \pi {R}^{2}$.


\begin{figure}[!htpb]
\centering
    \includegraphics[width=5cm]{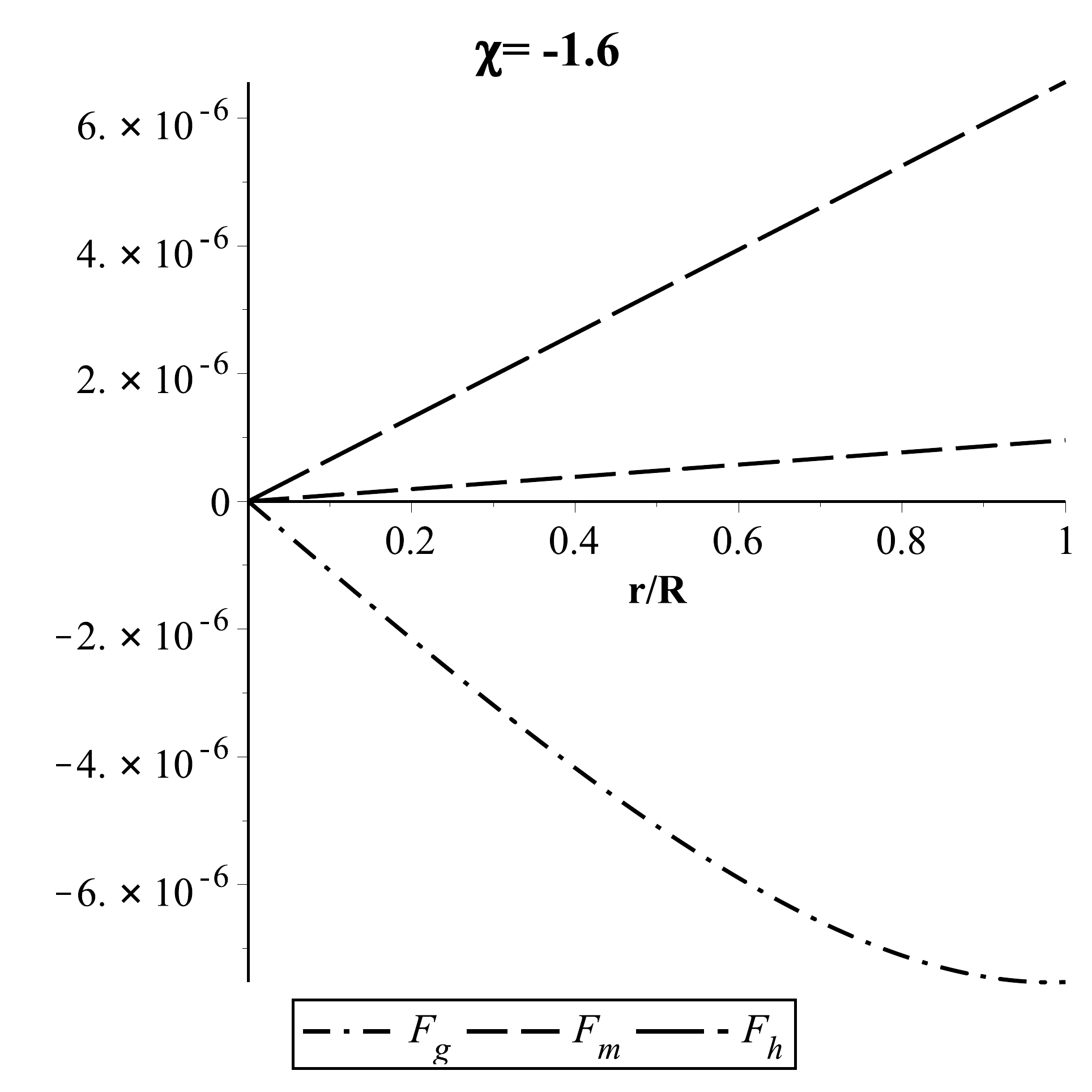}
    \includegraphics[width=5cm]{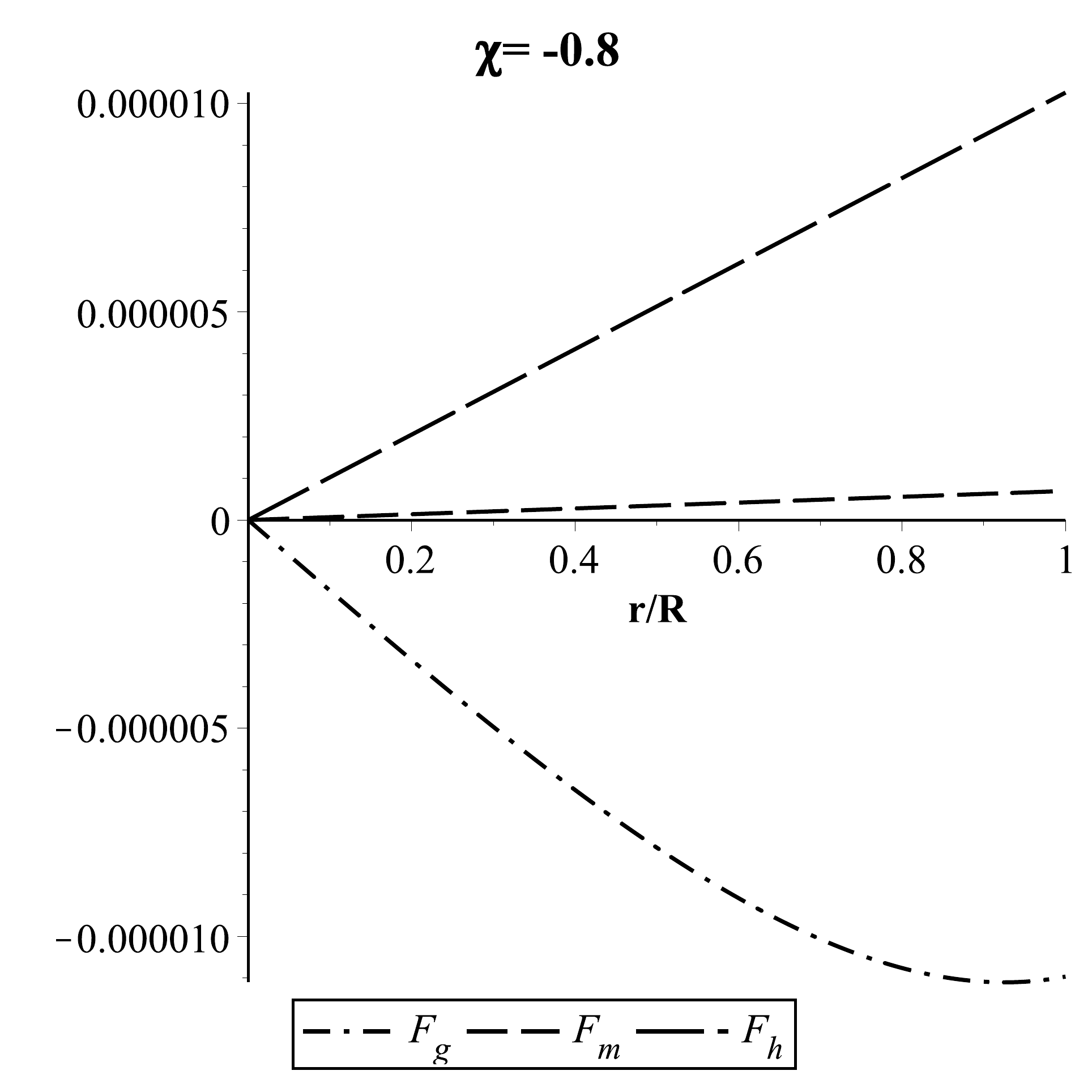}
    \includegraphics[width=5cm]{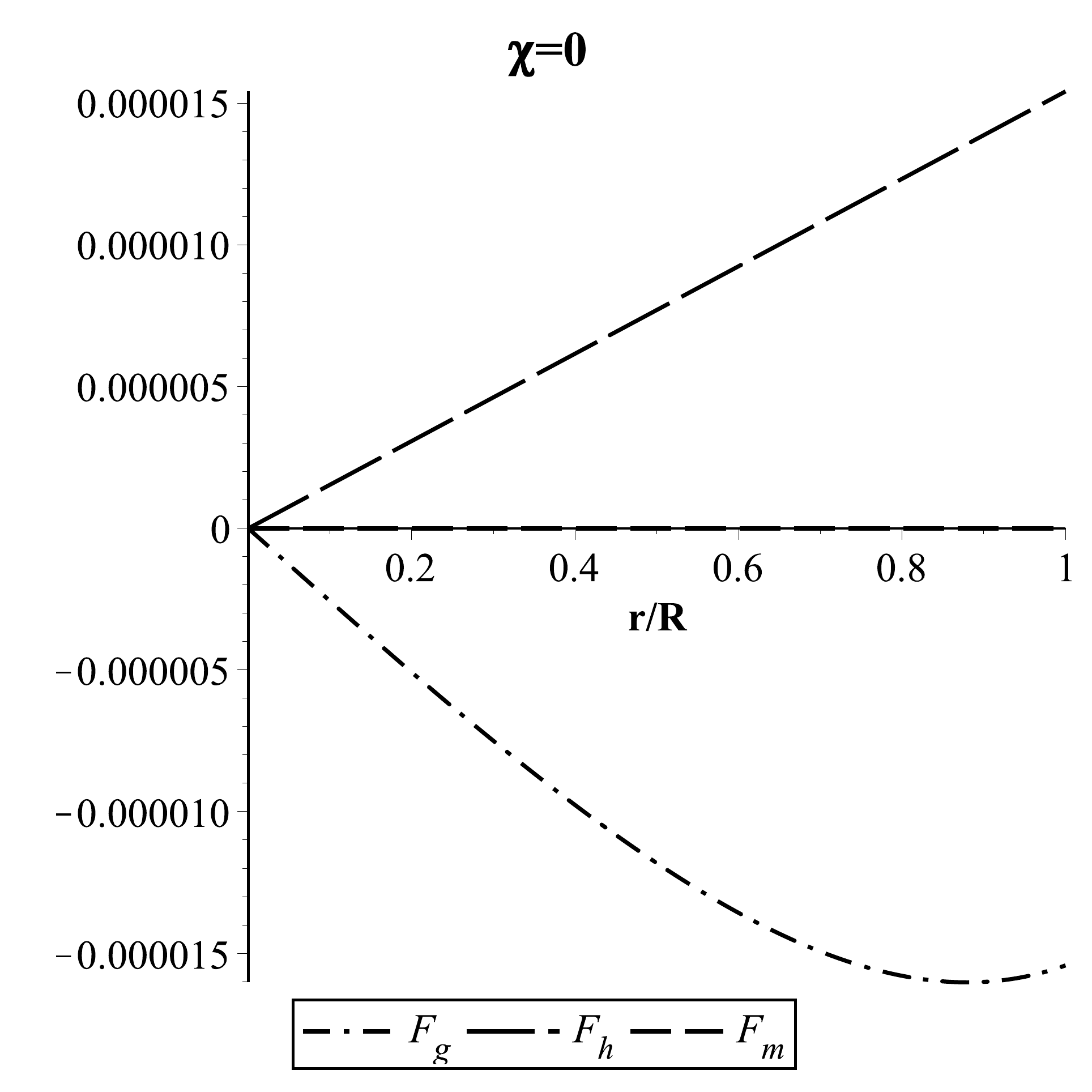}    
    \includegraphics[width=5cm]{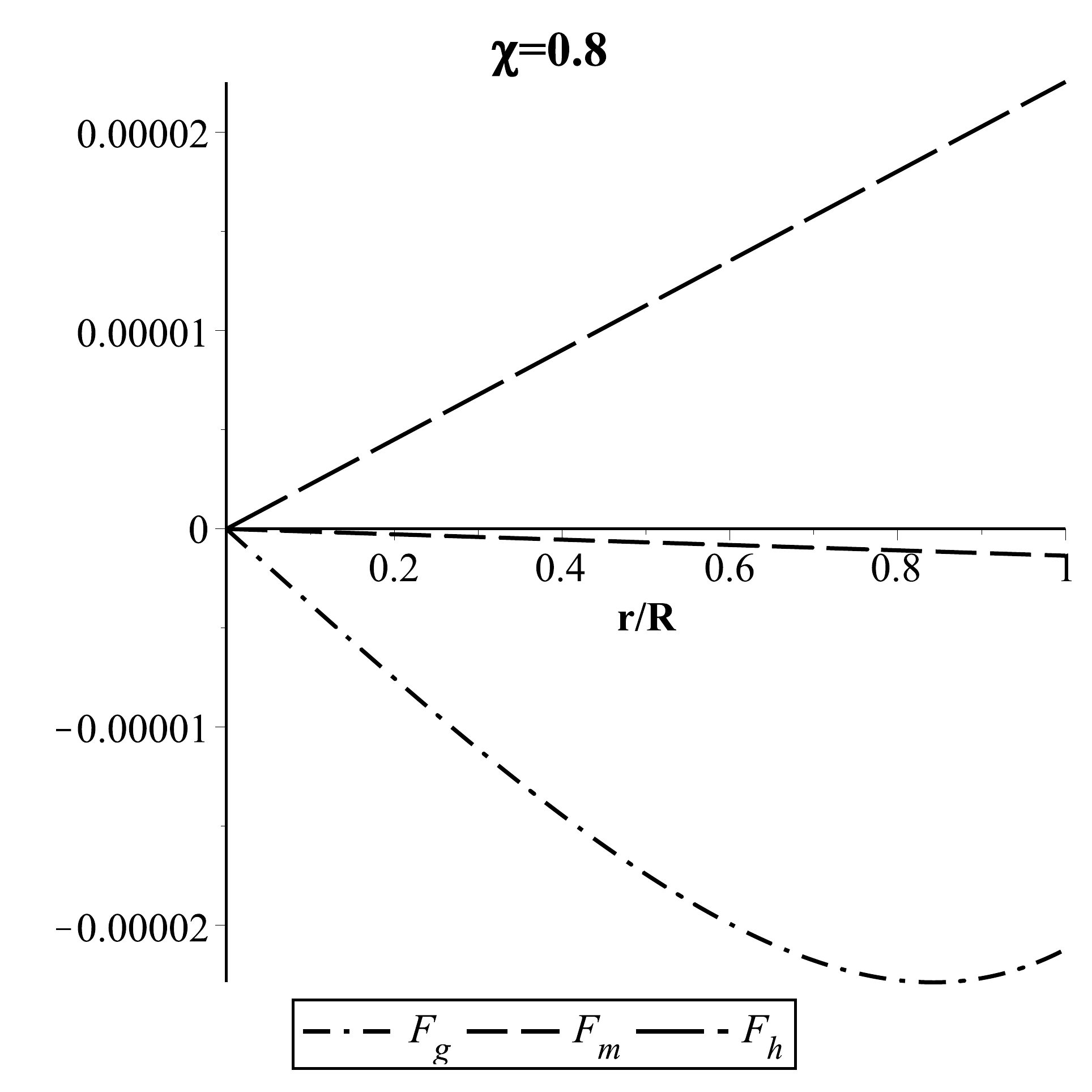}
   \includegraphics[width=5cm]{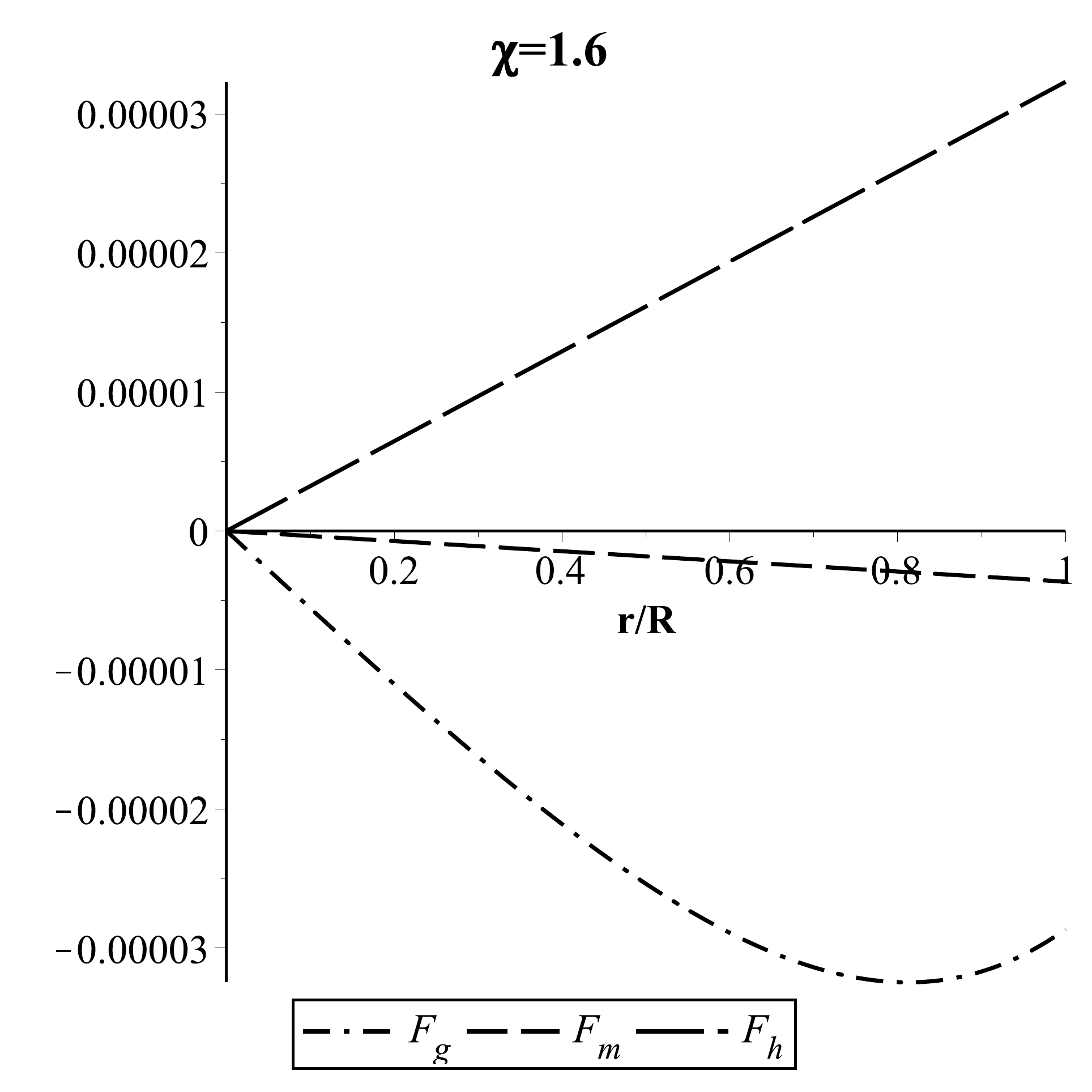}   
   \caption{Variation of forces with the radial coordinate $r/R$ for the strange star candidate $LMC~X-4$ due to different chosen values of $\chi$.} \label{Fig7}
\end{figure}


Figure~\ref{Fig7} features that the equilibrium of forces are achieved and this supports stability of the system. For $\chi>0$ the combined effect of $F_g$ and $F_m$ is counterbalanced by $F_h$ and for $\chi<0$ the effect of $F_g$ is counterbalanced by the combined effect of $F_h$ and $F_m$. For $\chi>0$ the force $F_m$ is attractive in nature and acts along the inward direction, however for $\chi<0$ we find $F_m$ is repulsive in nature and acts along the outward direction. For $\chi=0$ the effect of $F_g$ is counterbalanced by $F_h$ and the standard result of GR is retrieved.

\subsubsection{Causality condition}\label{subsubsec4.4.2}
According to the causality condition for a physically stable stellar model, the square of the sound speed~$\left({v^2_s}=\frac{d{p^{eff}}}{d{{\rho}^{eff}}} \right)$ should lie within the limit $\left[0,1\right]$, i.e., $0\leq {v^2_s} \leq 1$.  In the present study square of the sound speed is given by
\begin{eqnarray}\label{4.4.2.1}
{v^2_s}={\frac {\pi }{3\pi +\chi}}.
\end{eqnarray}


\begin{figure}[!htpb]
\centering
\includegraphics[width=6cm]{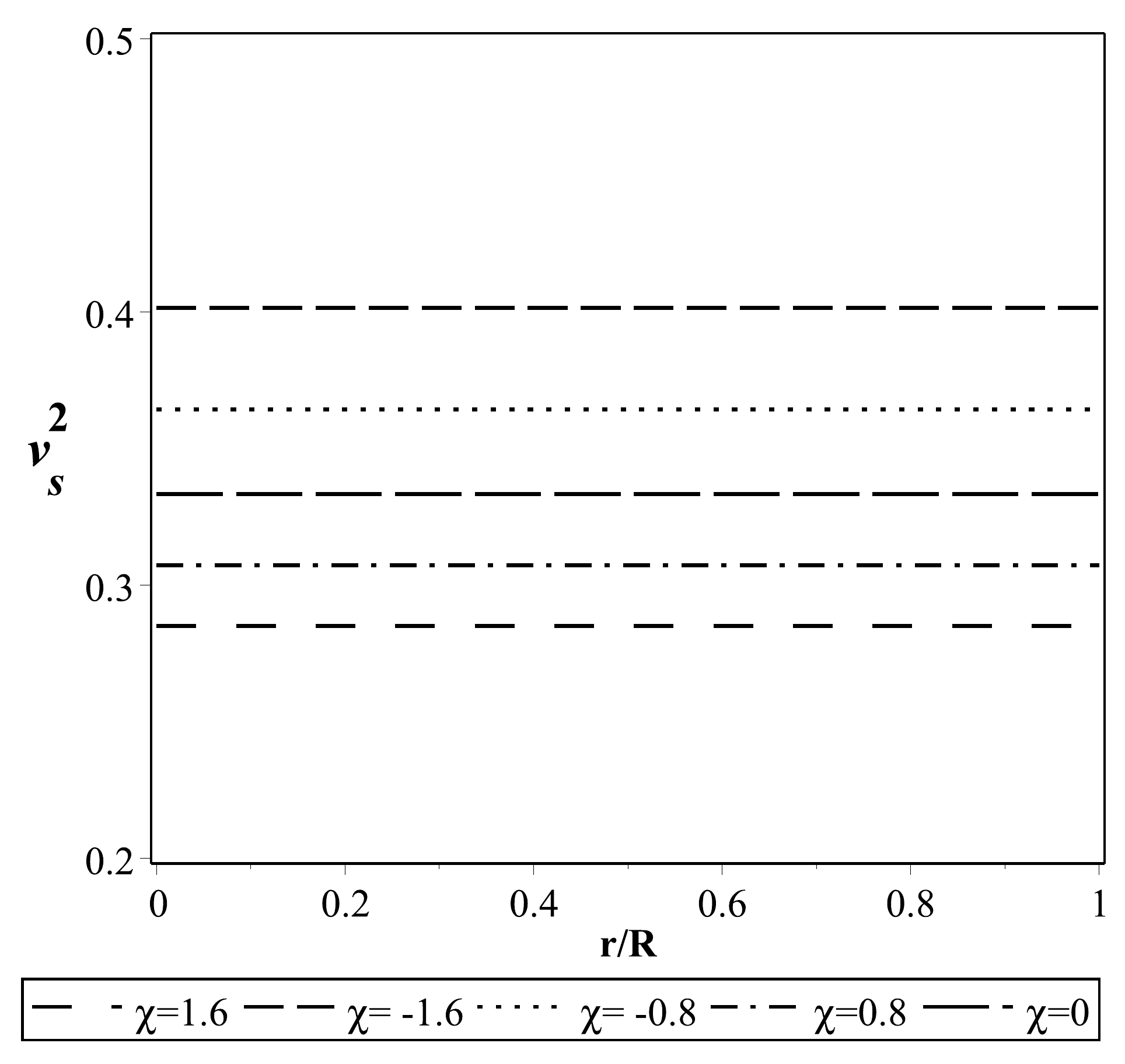}
\caption{Variation of ${v^2_s}$ with the radial coordinate $r/R$ for the strange star candidate $LMC~X-4$.} \label{Fig8}
\end{figure}


Figure~\ref{Fig8} features that for $\chi >0$ the value of ${v^2_s}$ decreases than the reference value of  ${v^2_s}$ in GR which is $\frac{1}{3}$ and for $\chi<0$ the value of ${v^2_s}$ increases than the value of ${v^2_s}$ as in $\chi=0$. For all the chosen values of $\chi$, we find that ${v^2_s}$ is well within the limit $\left[0,1\right]$. Hence, our model is consistent with the causality condition and is stable.

\subsubsection{Adiabatic Index}\label{subsubsec4.4.3}
The adiabatic index characterizes stiffness of the EOS for a given density and it can be used to study the stability of both relativistic and non-relativistic compact stars. After the pioneering work by Chandrasekhar~\cite{Chandrasekhar1964}, later on many scientists~\cite{Hillebrandt1976,Horvat2010,Doneva2012,Silva2015} have studied dynamical stability of the stellar system against an infinitesimal radial adiabatic perturbation. Heintzmann and Hillebrandt~\cite{Heintzmann1975} predicted that the adiabatic index should exceed $\frac{4}{3}$ inside a dynamically stable stellar system. For our model adiabatic index $\Gamma$ reads as
\begin{eqnarray}
& \qquad\hspace{-7cm} \Gamma=\frac{{p^{eff}}+{{\rho}^{eff}}}{{p^{eff}}}\,\frac{d{p^{eff}}}{d{{\rho}^{eff}}}=\frac{{p^{eff}}+{{\rho}^{eff}}}{{p^{eff}}}\,v^2_s\nonumber \\
& \qquad =\frac{14 \left[  \left( \pi +\frac{3\chi}{14}\right)  \left( \pi +\frac{\chi}{2}\right) B{
R}^{5}-{\frac {10}{7}}\lambda_{{1}}{r}^{2}{R}^{3}-{\frac {15}{56}}M\pi {R}^{2}+{\frac {15}{56}}M\pi {r}^{2} \right]  \left( \pi +\frac{\chi}{4}\right)}{15\left( \lambda_{{1}}{R}^{3}-\frac{3}{16}\,M\pi  \right)  \left( R-r \right) 
 \left( R+r \right)  \left( \pi +\frac{\chi}{3}\right)}. \label{4.4.3.1}
\end{eqnarray}


\begin{figure}[!htp]
\centering
    \includegraphics[width=6cm]{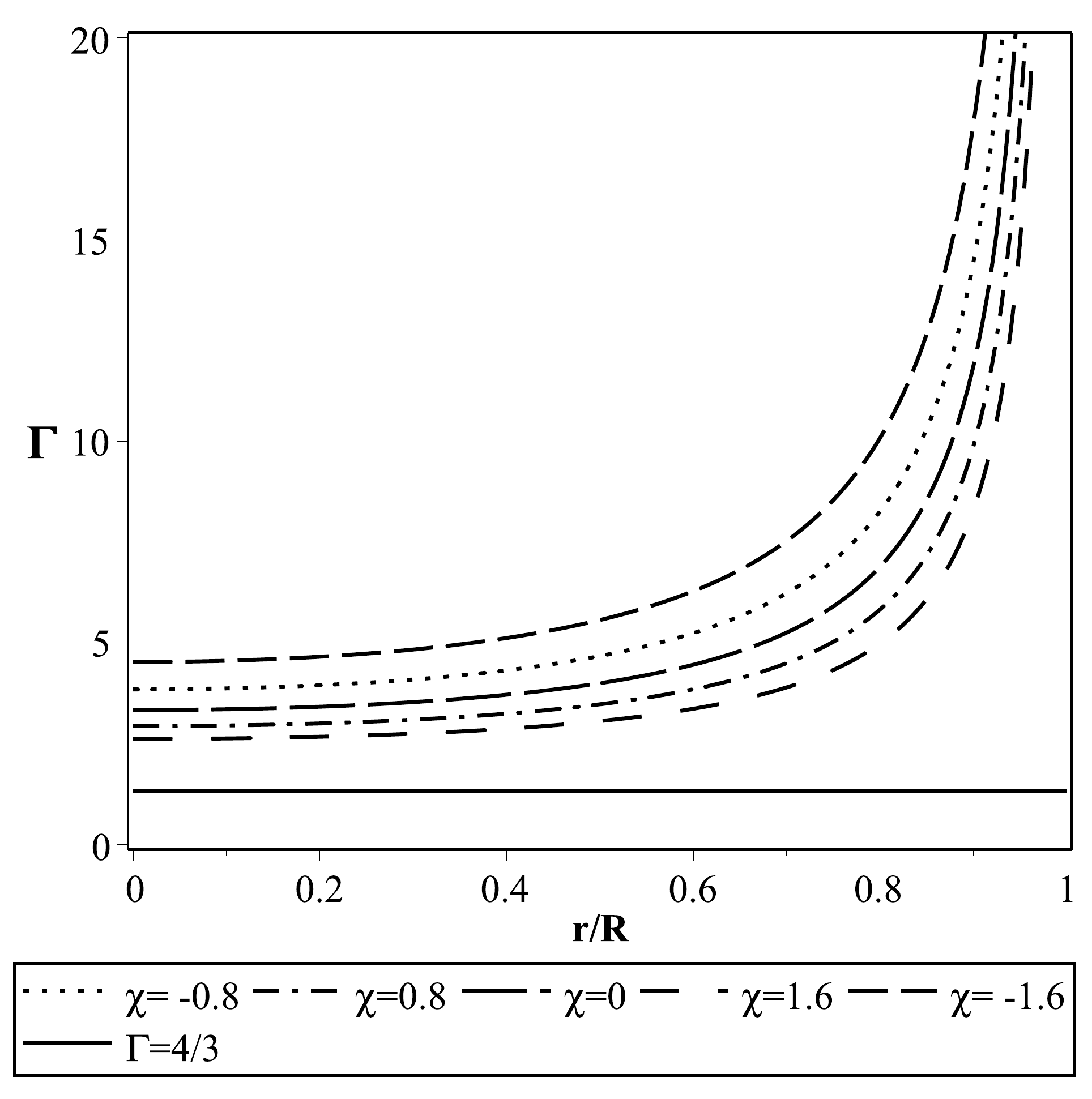}
    \caption{Variation of adiabatic indeces $\Gamma$ with the radial coordinate $r/R$ for the strange star candidate $LMC~X-4$.} \label{Fig9}
\end{figure}


The variation of $\Gamma$ is shown in figure~\ref{Fig9} and it reveals that presnt model is completely stable against an infinitesimal radial adiabatic perturbation for all the selected values of $\chi$ .

\section{Discussion and conclusion}\label{sec5}
In the present article following Harko et al.~\cite{harko2011} we choose a Lagrangian density as a simplified linear function of $f\left(R,\mathcal{T}\right)$ given as $f\left(R,\mathcal{T}\right)=R+2\chi T$, where $R$ is the Ricci scalar and $\mathcal{T}$ is the trace of the energy momentum tensor $T_{\mu\nu}$. Equation~(\ref{1.6}) clearly suggests that the system is not conserved in the point of view of GR as $\nabla^{\mu}T_{\mu\nu} \neq 0$. Again, eq.~(\ref{1.5}) indicates that the matter part is made of not only one type of fluid, which in our case is supposed to be the quark matter, but also a second type of fluid due to the effect of the coupling of matter and geometry. If we consider the energy-momentum tensor due to the effective fluid then the stellar system is conserved as shown in eq.~(\ref{1.7}) and field equations become  equivalent to the standard form of the Einstein gravity. For the detailed study, the work by Chakraborty~\cite{SC2013} (and the references therein) can be consulted where it has been explicitly discussed about the origin and nature of the second type of fluid due to $f\left(R,\mathcal{T}\right)$ gravity. However, in the present article our aim has been to study a model of strange stars under the framework of $f\left(R,\mathcal{T}\right)$ gravity and examine the stability as well as the physical properties of the ultra-dense compact stellar system.

Considering the simplified MIT bag EOS for quark matter, we have solved Einstein field equations using the observed values of the mass of some of the strange star candidates. To feature different properties and results of the solutions for the chosen parametric values of $\chi$, we considered $LMC~X-4$ as a representative of the strange star candidates. We choose the values of $\chi$ as $\chi=-1.6,~-0.8,~0,~0.8$~and$~1.6$. 

Variation of the metric potentials with respect to the radial coordinate are shown in figure~\ref{Fig1} which suggests that our system is free from any sort of physical or geometrical singularity. In figure~\ref{Fig2} variation of ${\rho}^{eff}$ and $p^{eff}$ with respect to the radial coordinate are presented. It shows that both the physical parameters are maximum at the center and decrease monotonically throughout the stellar system to become minimum at the surface. Our model is also consistent with all the energy conditions for all the chosen values of $\chi$ and they are shown in figure~\ref{Fig3}. Further, in figure~\ref{Fig4} we have presented the normalized total mass~$M$ versus the radius~$R$ relationship due to chosen values of $\chi$ for a specific value of bag constant given as $B=83~MeV/{fm}^{3}$~\cite{Rahaman2014}. We find that the maximum mass points (as indicated by the solid circles) and their radii decrease gradually with the increasing value of $\chi>0$ and they increase with the decreasing value of $\chi<0$. For example, when $\chi=-1.6$ the maximum mass and the corresponding radius increase to $30.49\%$ and $28.92\%$ than the respective values at $\chi=0$ and become $3.51~{M_{\odot}}$ and $12.63~km$, respectively, whereas the maximum mass and the corresponding radius decrease to $19.63\%$ and $18.8\%$ than the respective values as found in GR and become $2.17~{M_{\odot}}$ and $7.955~km$, respectively for $\chi=1.6$. 

In figure~\ref{Fig5} we have also shown the variation of the normalized total mass~$M$ and $R$ with respect to the central density of the effective fluid distribution (${{\rho}^{eff}_c}$). We find that with $\chi=-1.6$ the maximum mass point $3.52~{M_{\odot}}$ is obtained for ${{\rho}^{eff}_c}=2.076~{{\rho}_{nuclear}}$, which is $40.57\%$ lower than the value as found in GR. For $\chi=1.6$ the maximum mass point, $2.17~{M_{\odot}}$ is achieved for ${{\rho}^{eff}_c}=5.349~{{\rho}_{nuclear}}$, which is $53.15\%$ higher than the value found in GR. Therefore, essentially it can be predicted that as the values of $\chi$ increase from $\chi=0$, the stellar configuration of a strange star get shrinked and the density inside the system increases gradually to produce an ultra-dense compact stellar system. On the other hand, as $\chi$ decreases from $\chi=0$, the stellar system becomes more massive and larger in size so that the density decreases gradually to produce a less dense compact stellar system.

We have studied the present system for the specific form of the density profile (eq. \ref{2.8}) which is advantageous to use, as it is providing singularity free solution at the center. Moreover, applying this known form of the density profile and MIT bag EOS we are able to provide exact solutions of the modified TOV equation (basic eq. (2.9) and hence modified eq. (4.6)) for the $f(R,\mathcal{T})$ theory of gravity,  without employing any complex method of the numerical analysis. It is to note that due to this assumption of the density profile we can present the typical mass~$M$ (normalized in solar mass) versus radius~$R$ relationship for the strange star candidates in figure~\ref{Fig4} which may encourage further study in this line for all other suitable density profiles.

For the sake of interest of a comparative study, in Table~\ref{Table 1} we have predicted values of different physical parameters for the strange star candidate $LMC~X-4$ due to parametric values of $\chi$. The results from Table~\ref{Table 1} strongly suggest the same conclusions which we already predicted from figures~\ref{Fig4} and~\ref{Fig5}. In figure~\ref{Fig6} we have shown the variation of compactification factor and redshift with the radial coordinate. For the chosen values of $\chi$ we find from Table~\ref{Table 1} that the redshift of $LMC~X-4$ lies within the range $0.23-0.36$. In Table~\ref{Table 2} we have presented values of different physical parameters for the different strange star candidates due to a specific values of $\chi$, viz. $\chi=-0.8$. 

In the present study high values of the surface and central densities, and the high redshift value confirm that the chosen compact stellar candidates are actually strange quark stars~\cite{Ruderman1972,Glendenning1997,Herjog2011}. Also, from both the Tables.~\ref{Table 1} and \ref{Table 2} we find that for all values of $\chi$ the values of $2M/R$ are well within the Buchdahl limit~\cite{Buchdahl1959}, i.e., $2M/R>8/9$. To discuss the stability of the system in details we have studied the modified TOV equation for $f\left(R,\mathcal{T}\right)$ gravity as given in eq.~(\ref{3.5}). Figure~\ref{Fig7} features that for both $\chi<0$ and $\chi>0$, the sum of all the forces are zero and the system attains equilibrium. It also shows an interesting fact that an extra force $F_m$ arises due to $f\left(R,\mathcal{T}\right)$ gravity. This factor $F_m$ is repulsive in nature and acts along the outward direction for $\chi<0$, however the force is attractive in nature which acts along the inward direction when $\chi>0$. The square of the sound speed ($v_s^2$) due to different values of $\chi$ is constant throughout the stellar configuration and well within the limit $\left[0,1\right]$, which confirms stability of the system. The variation of $v_s^2$ with respect to the radial coordinate is shown in figure~\ref{Fig8}. Also, from figure~\ref{Fig9} we find that the adiabatic indices ($\Gamma$) for the chosen values of $\chi$ are greater than $4/3$ and thus provide stability of the system against an infinitesimal radial adiabatic perturbation.

In a nutshell, in the present study by providing an exact solution of the Einstein field equations and examining different physical tests under $f\left(R,\mathcal{T}\right)$ theory of gravity we have presented a stable stellar model suitable for the strange star candidates.

\section*{Acknowledgments}
SR and FR are thankful to the Inter-University Centre for
Astronomy and Astrophysics (IUCAA), Pune, India for providing
Visiting Associateship under which a part of this work was carried
out. SR is also thankful to the authority of The Institute of
Mathematical Sciences, Chennai, India for providing all types
of working facility and hospitality under the Associateship scheme. 
FR is also thankful to DST-SERB (EMR/2016/000193), Govt. of India 
for providing financial support. A part of this work was completed 
while D.D. was visiting IUCAA and the author gratefully 
acknowledges the warm hospitality and facilities at the library there. 
We all are thankful to the anonymous referee for several pertinent suggestions 
which have enabled us to modify the manuscript substantially.

\end{document}